\documentclass{kapedbk}
\usepackage{graphicx}
\normallatexbib
\begin{document}
\articletitle{Optical signatures of electron \\
correlations in the cuprates}
\author{D. van der Marel}
\affil{Laboratory of Solid State Physics\\
 Materials Science Centre, University of Groningen\\
 Nijenborgh 4, 9747 AG Groningen, The Netherlands}\email{d.van.der.marel@phys.rug.nl}
\begin{abstract}
The f-sum rule is introduced and its applications to electronic
and vibrational modes are discussed. A related integral over the
intra-band part of $\sigma(\omega)$ which is also valid for
correlated electrons, becomes just the kinetic energy if the only
hopping os between nearest neighbor sites. A summary is given of
additional sum rule expressions for the optical conductivity and
the dielectric function, including expressions for the first and
second moment of the optical conductivity, and a relation between
the Coulomb energy and the energy loss function. It is
shown from various examples, that the optical spectra of high
T$_c$ materials along the c-axis and in the ab-plane direction can
be used to study the kinetic energy change due to the appearance
of superconductivity. The results show, that the pairing mechanism
is highly unconventional, and mostly associated with a lowering of
kinetic energy parallel to the planes when pairs are formed.
\end{abstract}

\section{Macroscopic electromagnetic fields in matter}
\subsection{Introduction}\label{sec:intro}
The response of a system of electrons to an externally applied
field is commonly indicated as the dielectric function, or
alternatively as the optical conductivity. The discussion in this
chapter is devoted to induced currents and fields which are
proportional to the external fields, the so-called linear
response. The dielectric and the optical conductivity can be
measured either using inelastic scattering of charged particles
for which usually electrons are used, or by measuring the
absorption of light, or the amplitude and/or phase of light
reflected or transmitted by a sample. The two cases of fast
particles and incident radiation involve different physics and
will be discussed separately.
\subsection{Reflection and refraction of electromagnetic waves}\label{sec:reflection}
Optical spectroscopy measures the reflection and refraction of a
beam of photons interacting with the solid. A rarely used
alternative is the use of bolometric techniques to measure the
absorption of photons directly. A variety of different
experimental geometries can be used, depending on the type of
sample under investigation, which can be a reflecting surface of a
thick crystal, a free standing thin film, or a thin film supported
by a substrate. Important factors influencing the type of analysis
are also the orientation of the crystal or film surface, the angle
of incidence of the ray of photons, and the polarization of the
light. In most cases only the amplitude of the reflected or
refracted light is measured, but sometimes the phase is measured,
or the phase difference between two incident rays with different
polarization as in ellipsometry. The task of relating the
intensity and/or phase of the reflected or refracted light to the
dielectric tensor inside the material boils down to solving the
Maxwell equations at the vacuum/sample, sample/substrate, {\em
etc.} interfaces. An example is the ratio of the reflection
coefficients ($R_p/R_s$)and phase differences ($\eta_p-\eta_s$) of
light rays with $p$ and $s$-polarization reflected on a
crystal/vacuum interface at an angle of incidence $\theta$.
These quantities which are measured
directly using ellipsometry
\begin{equation}
 e^{i(\eta_p-\eta_s)}\sqrt{\frac{R_p}{R_s}} =
           \frac{\sin\theta\tan\theta-\sqrt{\epsilon-\sin^2\theta}}
                {\sin\theta\tan\theta+\sqrt{\epsilon-\sin^2\theta}}
 \label{ellips}
\end{equation}
The real and imaginary part of the dielectric constant can be
calculated from such a measurement with the aid of Eq.
\ref{ellips}. In contrast to a beam of charged particles, the
electric field of a plane electromagnetic wave is transverse to
the photon momentum. The dielectric tensor elements which can be
measured in an optical experiment are therefore transverse to the
direction of propagation of the electromagnetic wave. In a typical
optical experiment the photon energy is below 6 eV. In vacuum the
photon wave number used in optical experiments is therefore $0.0005
\AA^{-1}$, or smaller, which is at least three orders of magnitude
below the Fermi momentum of electrons in a solid. For this reason
it is usually said that optical spectroscopy measures the
transverse dielectric constant or the optical conductivity at zero
momentum. The optical conductivity tensor expresses the  current
response to an electrical field
\begin{equation}
 \vec{j}(\vec{r},t) = \int d^3\vec{r}' \int dt'
 \sigma(\vec{r},\vec{r}',t-t') \vec{E}(\vec{r}',t')
\end{equation}
From the Maxwell equations it can be shown that for polarization
transverse to the propagation of an electromagnetic wave
$d\vec{E}/dt = d\vec{D}/dt + 4\pi \vec{j}$. If the sample has
translational invariance, the optical conductivity tensor has a
diagonal representation in $k$-space. Due to the fact that the
translational symmetries of a crystalline solid are restricted to
a discrete space group, $k$ is limited to the first Brillouin
zone. Consequently, as shown by Hanke and Sham\cite{hanke75},
the $k$-space representation of the dielectric tensor becomes
a matrix in reciprocal space
\begin{equation}
 \sigma(\vec{G},\omega)_{\vec{G},\vec{G}'}=
 \int d^2\vec{r}\int d^3 \vec{r}' \int dt
   e^{i(\vec{q}+\vec{G})\cdot\vec{r}}
   e^{-i(\vec{q}+\vec{G}')\cdot\vec{r}'}
   e^{i\omega t}
   \sigma(\vec{r},\vec{r}',t)
\end{equation}
The dependence of $\sigma(\vec{q},\omega)_{\vec{G},\vec{G}'}$ on
the reciprocal lattice vectors $\vec{G}$, $\vec{G}'$ reflects,
that the local fields can have strong variations in direction and
magnitude on the length scale of a unit cell. Yet due to the long
wavelength of the external light rays the Fresnel equations
involve only $\vec{G}=\vec{G}'=0$. Usually in texts on optical
properties the only optical tensor elements considered have
$\vec{G}=\vec{G}'=0$, and in this chapter we will do the same.

Inside a solid the wavelength of the electromagnetic rays can be
much shorter than that of a ray with the same frequency travelling in
vacuum. Although in this chapter we will not encounter experiments
where the finite momentum of the photon plays an important role,
we should keep in mind that in principle the photon momentum is
non-zero and can have a non-trivial effect on the optical spectra.
In particular it may corrupt Kramers-Kronig relations, which is
just one out of several reasons why spectroscopic ellipsometry
should be favored.

\subsection{Inelastic scattering of charged particles}
When a fast charged particle, moving at a velocity $\vec{v}_e$,
interacts weakly with a solid, it may recoil inelastically by
transferring part of its momentum, $\hbar\vec{q}$ and its energy,
$\hbar\omega$ to the solid. The fast electron behaves like a test
charge of frequency $\omega=\vec{q}\cdot\vec{v}_e$, which
corresponds to a dielectric displacement field, $D(\vec{r},t) = e
q^{-2} \exp{(i\vec{q}\cdot\vec{r}-i\omega t)}$. The dielectric
displacement of the external charges may be characterized by a
density fluctuation, which has no field component transverse to
the wave. $D(\vec{r},t)$ is therefore a purely longitudinal field.
In a solid mixing of transverse and longitudinal modes occurs
whenever fields propagate in a direction which is not a high
symmetry direction of the crystal. However, in the long wavelength
limit the dielectric properties can be described by only three
tensor elements which correspond to the three optical axes of the
crystal. Since along these directions no mixing between
longitudinal and transverse response occurs, we will consider the
situation in this chapter where the fields and their propagation
vector point along the optical axis. Inside a material the
dielectric displacement is screened by the response of the matter
particles, resulting in the screened field $E(\vec{r},t)$  inside
the solid\cite{hanke75}.
\begin{equation}
 \vec{E}(\vec{r},t) = \int d^3\vec{r}' \int dt'
  \epsilon^{-1}(\vec{r},\vec{r}',t-t') \vec{D}(\vec{r}',t')
\end{equation}
For the same reasons as for the optical conductivity the
$k$-space representation of the dielectric tensor becomes a
tensor in reciprocal space
\begin{equation}
   \epsilon^{-1}(\vec{q}+\vec{G}',\vec{q}+\vec{G},\omega) =
   \int d^2\vec{r}\int d^3 \vec{r}' \int dt
   e^{i(\vec{q}+\vec{G}')\cdot\vec{r}}
   e^{-i(\vec{q}+\vec{G})\cdot\vec{r}'}
   e^{i\omega t}
   \epsilon^{-1}(\vec{r},\vec{r}',t)
\end{equation}
where $\vec{G}$ and $\vec{G}'$ denote reciprocal lattice vectors. The relation
between the dielectric displacement and the electric field is
\begin{equation}
   \vec{E}(\vec{q}+\vec{G}',\omega) = \sum_{\vec{G}}
    \epsilon^{-1}(\vec{q}+\vec{G}',\vec{q}+\vec{G},\omega)
   \vec{D}(\vec{q}+\vec{G},\omega)
\end{equation}
The macroscopic dielectric constant, which measures the
macroscopic response to a macroscopic perturbation, {\em i.e.} for vanishingly
small $\vec{q}$, is given by\cite{hanke75}
\begin{equation}
\epsilon(\omega)=\lim_{q\rightarrow 0} \frac{1}{\epsilon^{-1}(\vec{q},\vec{q},\omega)}
\end{equation}
where it is important, that in this expression first the matrix
$\epsilon(\vec{q}+\vec{G}',\vec{q}+\vec{G},\omega)$ has to be
inverted in reciprocal space, and in the next step the
$(\vec{G}=0,\vec{G}'=0)$ matrix element is taken of the inverted
matrix\cite{hanke75}. Energy loss spectroscopy using charged
particles can be used to measure the dielectric response as a
function of both frequency and momentum. This technique provides
the longitudinal dielectric function, {\em i.e.} the response to a
dielectric displacement field which is parallel to the transferred
momentum $\vec{q}$. The probability per unit time that a fast
electron transfers momentum $\vec{q}$ and energy $\hbar\omega$ to
the electrons was derived by Nozi\`eres and
Pines\cite{nozieres58,nozieres59} for a fully translational
invariant 'jellium' of interacting electrons
\begin{equation}
 P(\vec{q},\omega) = \frac{8\pi e^2}{|q|^2}
 \mbox{Im}\left\{\frac{-1}{\epsilon(\vec{q},\omega)}\right\}
 \label{charged particles}
\end{equation}
where $e$ is the elementary charge.
\subsection{Relation between $\sigma(\omega)$ and $\epsilon(\omega)$}
We close this introduction by remarking, that for electromagnetic
fields propagating at a long wavelength the two responses,
longitudinal and transverse, although different at any nonzero
wave vector, are very closely related.
We will take advantage of this fact when later in this chapter
we extract the energy loss function for $q\approx 0$ from optical data.
According to Maxwell's equations for $q\rightarrow 0$ the uniform
current density is just the time derivative of the uniform
dipole field, hence $4\pi j=i\omega (E-D)$. Consequently
for $q\rightarrow 0$ the conductivity and the dielectric function
are related in the following way
\begin{equation}
 \epsilon(0,\omega) = 1 + \frac{4\pi i}{\omega} \sigma(0,\omega)
 \label{epsilonsigma}
\end{equation}
Throughout this chapter we will use this identity repeatedly.

\section{Interaction of light with matter}
\subsection{The optical conductivity}
Let us now turn to the discussion of the microscopic properties of
the optical conductivity function. The full Hamiltonian describing
the electrons and their interactions is
\begin{eqnarray}
 H &=& \sum_{k\sigma} \frac{\hbar^2 k^2}{2m} c^{\dagger}_{k\sigma}c_{k\sigma}
 + \sum_{G}U_G \hat{\rho}_{-G} + \frac{1}{2}\sum_{k}V_{k}\hat{\rho}_{\vec{k}}\hat{\rho}_{-\vec{k}}\\
 \hat{\rho}_{\vec{k}}&=&\sum_{p,\sigma}c^{\dagger}_{p,\sigma} c_{p+k,\sigma}
 \label{fullhamiltonian}
\end{eqnarray}
In this expression the symbol $c^{\dagger}_{p,\sigma}$ creates a plane
wave of momentum $\hbar \vec{p}$ and spin-quantum number $\sigma$,
$U_G$ represents the potential landscape due to the crystal
environment. The third term is a model electron-electron interaction
Hamiltonian, representing all electron-phonon mediated and Coulomb interactions,
where $ \hat{\rho}_{\vec{k}}$ is the $k$'th Fourier component of the
density operator. In addition to the direct Coulomb interaction, various
other contributions may be relevant, such as direct exchange terms. As a result the
spin- and momentum dependence of the total interaction can have a more complex
form than the above model Hamiltonian. Relevant for the subsequent
discussion is only, that the interaction term commutes with
the current operator. The quantum mechanical expression for the current
operator is
\begin{eqnarray}
 \vec{j}_{\vec{q}}&=&\sum_{p,\sigma}\frac{e\hbar\vec{p}}{m}
  c^{\dagger}_{p-q/2,\sigma} c_{p+q/2,\sigma}
 \label{rho&jq}
\end{eqnarray}
The current and density operators are symmetric in k-space, satisfying
$\hat{\rho}^{\dagger}_k = \hat{\rho}_{-k}$ and $j^{\dagger}_k=j_{-k}$.
In coordinate space the representations of the density and the current are
\begin{eqnarray}
 \hat{n}(\vec{r})&=&\frac{1}{V}\sum_{q}e^{i\vec{q}\cdot\vec{r}}\hat{\rho}_{\vec{q}}\\
 \vec{j}(\vec{r})&=&\frac{e}{V}\sum_{q}e^{i\vec{q}\cdot\vec{r}}\vec{j}_{\vec{q}}
 \label{rho&jr}
\end{eqnarray}
It is easy to verify, that together $\hat{n}(\vec{r})$ and $\vec{j}(\vec{r})$
satisfy the continuity equation
$ie\hbar^{-1}[\hat{n}(\vec{r}),H] +\nabla\cdot \vec{j}(\vec{r}) =0$.

Let us now consider a many-body system with eigen-states
$|m\rangle$ and corresponding energies $E_{m}$. For such a system
the microscopic expression for the optical conductivity has been
explained by A.J. Millis in Chapter 6. The result for
finite $\vec{q}$ was derived in 'The theory of quantum liquids'
part I, by Nozi\`eres and Pines (equation 4.163). For brevity of
notation we represent the matrix elements of the current operators
as
\begin{equation}
      j_{\alpha,q}^{nm} \equiv \langle n|j_{\alpha,q}|m\rangle
 \label{jmn}
\end{equation}
With the help of these matrix elements, and with the definition
$\hbar\omega_{mn}= E_{m}-E_{n}$
the expression for the optical conductivity is
\begin{equation}
 \sigma_{\alpha,\alpha}(\vec{q},\omega)=  \frac{i e^2 N}{mV\omega}
  + \sum_{n, m \neq n} \frac{i e^{\beta(\Omega-E_n)}}{V\omega}
  \left[
           \frac{ j_{\alpha,q}^{nm}j_{\alpha,-q}^{mn} }
                {\omega-\omega_{mn}+i\eta}
         -
           \frac{ j_{\alpha,-q}^{nm}j_{\alpha,q}^{mn} }
                {\omega+\omega_{mn}+i\eta}
  \right]
 \label{sigmamillis}
\end{equation}
Here $N$ is the number of electrons, $V$ the volume, $m$ the electron mass,
$q_e$ the elementary charge, $\Omega$ is the thermodynamic potential, $\beta=1/k_BT$ and
$\eta$ is an infinitesimally small positive number. In principle
in the calculation of Eq. \ref{sigmamillis}
terms may occur under the summation for which $\omega_{mn}=0$. As
$\omega_{mn}$ occurs in the denominator of this expression, these zeros
should be cancelled exactly by zeros of the current matrix elements, which
poses a special mathematical challenge.

In Eq. \ref{sigmamillis}
$\sigma(\omega)$ is represented by two separate terms, a
$\delta$-function for $\omega=0$ and a summation over
excited many-body eigen-states. The $\delta$-function is a diamagnetic
contribution of {\em all} electrons in the system, the presence of
which is a consequence of the gauge invariant treatment of the
optical conductivity, as explained by Millis in chapter 6.
The presence of this term is at first glance rather confusing,
since left by itself this $\delta$-function would imply that all materials
(including diamond) are ideal conductors! However,
the second term has, besides a series of poles
corresponding to the optical transitions, also a pole
for $\omega=0$, corresponding to a {\em negative} $\delta$-function
of Re$\sigma(\omega)$. It turns out, that for all materials except
ideal conductors this $\delta$-function compensates exactly the first
(diamagnetic) term of Eq. \ref{sigmamillis}. This exact
compensation is a consequence of the relation
\cite{proof1}
\begin{equation}
     \mbox{For every $n$:   }
     \sum_{m \neq n} \frac{j_{\alpha,q}^{nm}j_{\alpha,-q}^{mn}}{\omega_{mn}}
      = \frac{Ne^2}{2m}
 \label{sum1}
\end{equation}
Experimentally truly 'ideal'
conductivity is only seen in superconductors. In ordinary conducting materials
the diamagnetic term broadens to a Lorentzian peak due to elastic and/or
inelastic scattering. The width of this peak is the inverse
life-time of the charge carriers. Often in the theoretical
literature the broadening is not important, and the Drude peak is counted
to the Dirac-function in the origin. The infrared properties of
superconductors are characterized by the presence of both a
purely reactive diamagnetic response, and a regular dissipative conductivity
\cite{footnotesuperconductor}. The sum of these contributions counts the partial
intra-band spectral weight which we discuss in section \ref{sec:ksum}.
With the help of Eq.\ref{sum1}, the diamagnetic term of Eq. \ref{sigmamillis} can
now be absorbed in the summation on the right-hand side
\begin{equation}
 \sigma_{\alpha,\beta}(\vec{q},\omega)=
  \frac{i}{V} \sum_{n, m \neq n}
   \frac{e^{\beta(\Omega-E_{n})}}{\omega_{mn}}
  \left\{ \frac{j_{\alpha,q}^{nm}j_{\alpha,-q}^{mn}}
        {\omega-\omega_{mn}+i\eta}
  +     \frac{j_{\alpha,-q}^{nm}j_{\alpha,q}^{mn}}
        {\omega+\omega_{mn}+i\eta} \right\}
 \label{sigmaA}
\end{equation}
As explained in section \ref{sec:reflection}, usually in optical
experiments one assumes $q\rightarrow 0$ in the expressions for
$\sigma(\omega)$. It is useful at this stage to introduce the
generalized plasma frequencies $\Omega_{mn}^2 = 8\pi
e^{\beta(\Omega-E_{n})} |j_{\alpha}^{nm}|^2
\omega_{mn}^{-1}V^{-1}$, with the help of which we obtain the
following compact expression for the optical conductivity tensor
\begin{equation}
 \sigma_{\alpha\alpha}(\omega)=
  \frac{i\omega}{4\pi}  \sum_{n, m \neq n}
    \frac{\Omega_{mn}^2} {\omega(\omega+i\gamma_{mn})-\omega_{mn}^2}
 \label{sigmaBB}
\end{equation}
Although formally the parameter $\gamma_{mn}$ is understood to
be an infinitesimally small positive number, a natural modification of
Eq.\ref{sigmaBB} consists of limiting the summation to a set of oscillators
representing the main optical transitions and inserting a finite value
for $\gamma_{mn}$, which in this case represents the inverse lifetime
of the corresponding excited state ({\em e.g.} calculated using Fermi's Golden Rule).
With this modification Eq.\ref{sigmaBB} is one of the most
commonly used phenomenological representations of the optical conductivity,
generally known as the Drude-Lorentz expression.

\subsection{The f-sum rule}
The expressions Eqs. \ref{sigmamillis}, \ref{sigmaA}, and \ref{sigmaBB} satisfy a
famous sum rule. This is obtained by first showing with
the help of Eq. \ref{sum1}, that for each $n$
\begin{equation}
 \sum_{m, m \neq n} \Omega_{mn}^2 \equiv  \frac{4\pi e^2 N}{mV}
              e^{\beta(\Omega-E_{n})}
 \label{sum2}
\end{equation}
Second, as a result of Cauchy's theorem in Eq.\ref{sigmaBB} the
integral over all (positive and negative) frequencies of
$\int$Re$\sigma(\omega)$ equals $\sum_{mn}  \Omega_{mn}^2 /4$. To
complete the derivation of the f-sum rule we also use that
$\sum_{n}e^{\beta(\Omega-E_{n})}=1$, which follows from the
definition of the thermodynamic potential. Then
\begin{equation}
 \int_{-\infty}^{\infty}
  \mbox{Re}\sigma(\omega) d\omega
   = \frac{\pi e^2 N}{m V}
 \label{sum3}
\end{equation}
is the f-sum rule, or Thomas-Reich-Kuhn rule. It is a cornerstone
for optical studies of materials, since it relates the integrated
optical conductivity directly to the density of charged objects,
and the absolute value of their charge and mass. It reflects the
fundamental property that also in strongly correlated matter the
number of electrons is conserved. Note that the right-hand side of
the f-sum rule is independent of the value of $\hbar$. Also the
f-sum rule applies to bosons and fermions alike. Because
Re$\sigma(\omega)$=Re$\sigma(-\omega)$ the sum rule is often
presented as an integral of the conductivity over positive
frequencies only. Superconductors present a special case, since
Re$\sigma(\omega)$ now has a $\delta$ function at $\omega=0$: Only
half of the spectral weight of this $\delta$-function should be
counted to the positive frequency side of the spectrum.

\subsection{Spectral weight of electrons and optical phonons}
The optical conductivity has contributions
from both electrons and nuclei because each of these
particles carries electrical charge. The integral over the optical
conductivity can then be extended to the summation over
all species of particles in the solid with mass $m_j$, and charge $q_j$
\begin{equation}
     \int_{-\infty}^{\infty}\mbox{Re}\sigma(\omega)d\omega
     = \sum_{j} \frac{\pi q_j^2  N_j}{m_j V}
 \label{fsum}
\end{equation}
Because the mass of
an electron is several orders of magnitude lower than the mass of a
proton, in many cases the contribution of the nuclei to the f-sum rule
is ignored in calculations of the integrated spectral weight of
metals. However, important exceptions exist where the phonon
contribution cannot be neglected, notably in the c-axis response
of cuprate high T$_c$ superconductors.
\begin{figure}[ht]
\centerline{\includegraphics[width=10cm,clip=true]{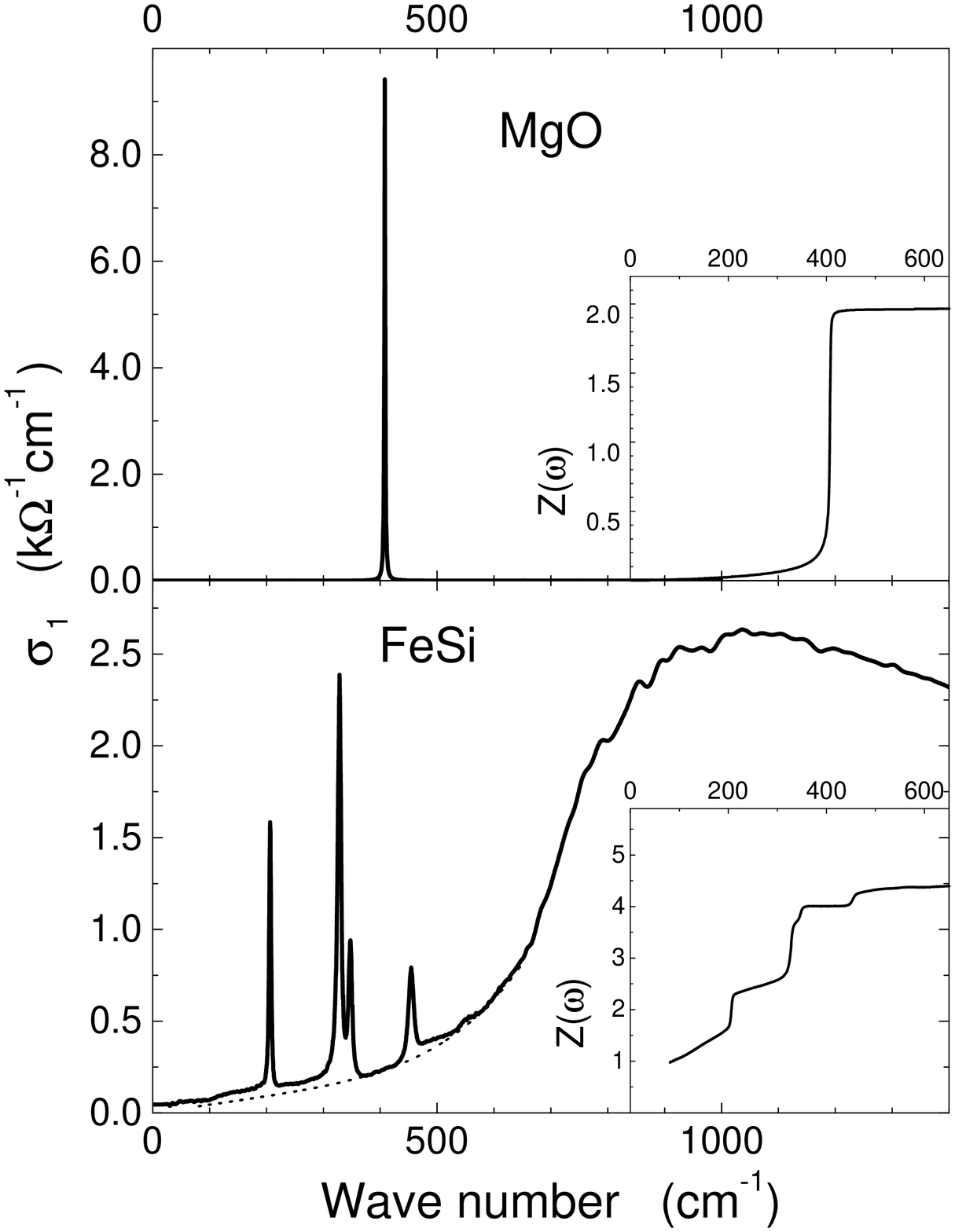}}
 \caption{\protect Optical conductivity of MgO (top panel) and FeSi at $T\!=\!4$ K
 (bottom panel). In the insets the function $Z(\omega)=8\mu (4\pi n_i e^2)^{-0.5}\left\{
 \int_0^{\omega}\sigma_{ph}(\omega')d\omega'\right\}^{0.5}$ is displayed. For FeSi the electronic
 background (dotted curve of the lower panel) was subtracted. For $\omega > 600 \mbox{cm}^{-1}$
 $Z(\omega)$  corresponds to
 transverse effective charge. Data from Ref. \cite{damascelli97,thesisandrea}.}
\label{fig:fsz}
\end{figure}
Although Eq.\ref{fsum} is completely general, in practice it cannot be applied to experimental spectra
directly. This is due to the fact that the contributions of all electrons and nuclei can only be obtained if the
conductivity can be measured sufficiently accurately up to infinite frequencies. In practice one always uses a
finite cut-off. Let us consider the example of an ionic insulator: If the integral is carried out for
frequencies including all the vibrational modes, but does not include any of the inter-band transitions, then the
degrees of freedom describing the motion of electrons relative to the ions is not counted. As a result the large
number of electrons and nuclei which typically form the ions are not counted as separate entities. Effectively
the ions behave as the only (composite) particles in such a case, and the right-hand side of Eq. \ref{fsum}
contains a summation over the ions in the solid. Application of Eq.\ref{fsum} provides the so-called transverse
effective charge, which for ionic insulators with a large insulating gap corresponds rather closely to the
actual charge of the ions. In the top panel of Fig. \ref{fig:fsz} this is illustrated with the infrared spectrum
of MgO. Indeed the transverse effective charge obtained from the sum rule is 1.99, in good agreement with the
formal charges of the Mg$^{2+}$ and O$^{2-}$ ions.

\begin{figure}[ht]
\centerline{\includegraphics[width=10cm,clip=true]{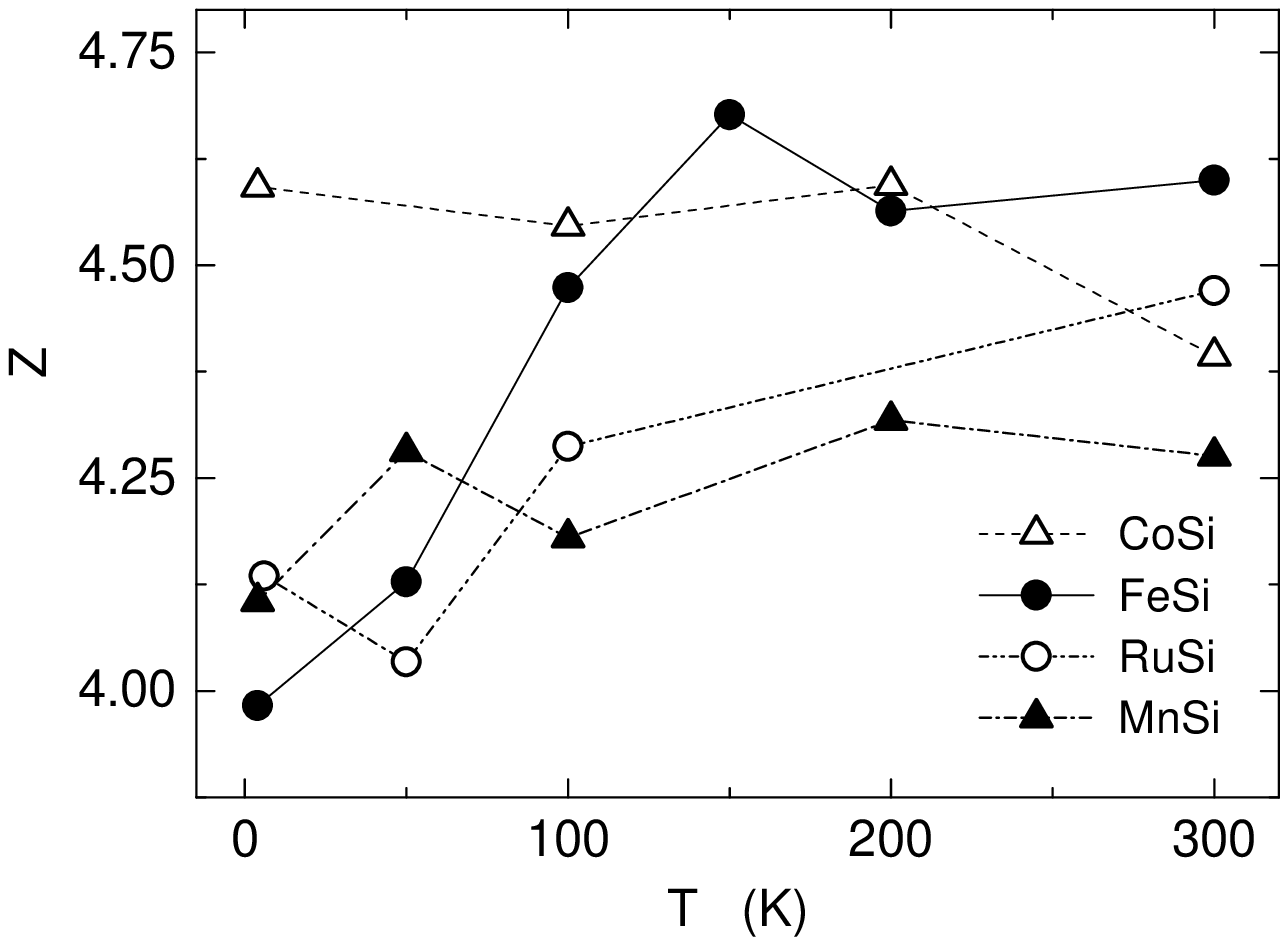}}
 \caption{\protect Temperature dependence of the transverse effective charge of Co-Si, Fe-Si, Ru-Si,
 and Mn-Si pairs, calculated from the oscillator strength of the optical phonons.
 Data from Refs. \cite{damascelli97}, \cite{thesisandrea} and \cite{verena98}. }
\label{fig:ztot}
\end{figure}

Because the mass of the ions is much higher than the free electron mass, the
corresponding spectral weight integrated over the vibrational part of the
spectrum is rather small. In a metal, even if optical phonons are present, usually the
spectral weight at low frequencies is completely dominated by the
electronic contributions due to the fact that the free electron
mass is much smaller than the nuclear mass. A widely spread
misconception is, that the screening of optical phonons in metals
leads to a smaller oscillator strength than in ionic insulators. The
opposite is true: Due to resonant coupling between vibrational
modes and electronic oscillators, the optical phonons in an
intermetallic compound often have much {\em more} spectral weight
than optical phonons in insulators. This 'charged phonon' effect
was formulated in an elegant way in 1977 by Rice, Lipari and
Str\"assler\cite{rice77}. They demonstrated, that under resonant conditions,
due to electron-phonon coupling, vibrational modes borrow oscillator
strength from electronic modes, which boosts the intensity of
the vibrational modes in the optical conductivity spectra.
This effect is now known to be common in many materials, for example in
TCNQ-salts,blue bronze, IV-VI narrow-gap semiconductors, FeSi and related
compounds, and the beta-phase of sodium vanadate
\cite{challener84,travaglini84,degiorgi90,degiorgi91,lucovsky73,thesisandrea,damascelli97,verena98,presura02}.
In  Figs. \ref{fig:fsz} and \ref{fig:ztot} the charged phonon
effect is illustrated using the examples of FeSi, MnSi, CoSi
and RuSi\cite{thesisandrea,damascelli97,verena98}, showing that the transverse
charge is between 4 and 5. These compounds are not ionic insulators,
because the TM and Si atoms have practically the same electro-negativities and
electron affinities. Instead the large transverse charge
of these compounds arises from the charged phonon effect predicted
by Rice. The strong temperature dependence of the transverse charge of FeSi
correlates with the gradual disappearance of the semiconductor gap as the temperature
is raised from 4 to 300 K.
\subsection{Partial spectral weight of the intra-band transitions}\label{sec:ksum}
Often there is a special interest in the spectral properties of
the charge carriers. The electrons are subject to the
periodic potential of the nuclei, resulting in an energy-momentum
dispersion which differs from free electrons. Often one takes
this dispersion relation as the starting point for models of
interacting electrons. The Coulomb interaction and other ({\em
e.g.} phonon mediated) interactions present the real theoretical
challenge. The total Hamiltonian describing the electrons
and their interactions is then
\begin{equation}
H=\sum_{k,\sigma} \epsilon_k c^{\dagger}_{k,\sigma}c_{k,\sigma} +
 \frac{1}{2}\sum_{kqp}V_{k}
 \sum_{\sigma\sigma'} c^{\dagger}_{p\sigma}c^{\dagger}_{q\sigma'}c_{q-k\sigma'}c_{p+k\sigma}
 \label{hamiltonian}
\end{equation}
The current operator is in this case
\begin{equation}
  \vec{j}_q=  \frac{e}{2}\sum_{p,\sigma}
  \left(v_{p+q/2}-v_{-p+q/2}\right) c^{\dagger}_{p-q/2,\sigma} c_{p+q/2,\sigma}
\end{equation}
where the $v_{k}\equiv\hbar^{-1}\partial\epsilon_k/\partial \vec{k}$ is the group velocity. The density operator
commutes with the interaction part of the Hamiltonian. This has an interesting and very useful consequence,
namely that a partial sum rule similar to the f-sum rule exists, which can be used to probe experimentally the
kinetic energy term of the Hamiltonian. This partial sum rule for integration of the intra-band conductivity,
$\sigma^{L}_{\alpha\alpha}(\omega)$, yields\cite{norman02}
\begin{eqnarray}
     \int_{-\infty}^{\infty}d\omega\mbox{Re}\sigma^{L}_{\alpha\alpha}(\omega)d\omega
     &=& \pi \frac{e^2}{V}  \sum_{k,\sigma}
       \langle c^{\dagger}_{k,\sigma}c_{k,\sigma}\rangle \frac{1}{m_k}\\
    \frac{1}{m_k} &=& \frac{1}{\hbar^2}
     \frac{\partial^2\epsilon(\vec{k})}{ \partial k_\alpha ^2}
 \label{fsum2}
\end{eqnarray}
Apparently the total spectral weight contained in the
inter-band transitions, $\sigma^{H}_{\alpha\alpha}(\omega)$, is exactly
\begin{equation}
        \int_{-\infty}^{\infty}d\omega\mbox{Re}\sigma^{H}_{\alpha\alpha}(\omega) =
        \pi e^2
        \left(\frac{n}{m}-\frac{1}{V}\sum_k \frac{n_k}{m_k} \right)
 \label{interband}
\end{equation}
In the limit where the interaction $V_k=0$, the occupation function $n_k$ in the above
summation is a step-function at the Fermi momentum. In this case the
summation over k becomes an integration over the Fermi volume with
$n_k$ set equal to 1. After applying Gauss's theorem we immediately
obtain the well-known Fermi surface integral formula
\begin{equation}
 \begin{array}{l}
     \omega_{p,\alpha}^2
     = g\int_{S_F} \frac{e^2}{\hbar} v_{\alpha}(\vec{a}) d a_{\alpha}
 \end{array}
 \label{fsum3}
\end{equation}
where $g$ is the spin degeneracy factor.
In the literature two limiting cases are most frequently considered: (i) the
free electron approximation, where $m_k=m_e$
is the free electron mass independent of the momentum
of the electron, and (ii) the nearest neighbor tight-binding limit. In the latter
case the dispersion is $\epsilon_k=-2t_x\cos k_xa_x-2t_y\cos k_ya_y-2t_z\cos k_za_z$, with
the effect that $1/m_k=-2t\hbar^{-2}a_{\alpha}^2\cos (k_{\alpha}a_{\alpha})$, and
\begin{equation}
 \begin{array}{l}
     \sum_{\alpha}\frac{\hbar^2}{a_{\alpha}^2 \pi e^2}\int_{-\omega_m}^{\omega_m}
     \mbox{Re}\sigma_{\alpha\alpha}(\omega)d\omega
       =  -\sum_{k,\sigma} n_k \epsilon_k = \langle - H_{kin} \rangle
 \end{array}
 \label{fsum4}
\end{equation}
where the integration should be carried out over all
transitions within this band, {\em including} the
$\delta$-function at $\omega=0$ in the superconducting state. The upper
limit of the integration is
formally represented by the upper limit $\omega_m$. In practice the
cutoff cannot always be sharply defined, because usually there is some
overlap between the region of transitions within the partially filled
band and the transitions between different bands. Hence in the
nearest neighbor tight-binding limit the f-sum provides the {\em
kinetic energy} contribution, which depends both on the number of
particles and the hopping parameter $t$\cite{maldague77}. This
relation was used by Baeriswyl {\em et al.} to show, using exact
results for one dimension, that the oscillator strength of optical
absorption is strongly suppressed if the on-site electron-electron
interactions (expressed by the Hubbard parameter $U$) are
increased\cite{baeriswyl86}. The same equation can also be applied
to superconductors, examples will be discussed later in this
chapter. In the case of a superconductor it is important to
realize, that the integration on the left-hand side of Eq.
\ref{fsum4} should also include the condensate $\delta$-function
at $\omega=0$. As the optical conductivity can only be measured
for $\omega > 0$, the spectral weight in the $\delta$-function has
to derived from a measurement of the imaginary part of
$\sigma(\omega)$, taking advantage of the fact that the real and
imaginary part of a $\delta$-function conductivity are of the form
\begin{displaymath}
  \sigma^{singular}(\omega)=\frac{2i\omega_{p,s}^2}{4\pi(\omega+i0^+)}
\end{displaymath}
The plasma frequency of the condensate, $\omega_{p,s}$, is
inversely proportional to the London penetration depth,
$\lambda(T) = c/ \omega_{p,s}(T)$ with $c$ the velocity of light.
In the literature\cite{basov99,chakravarty99} the
$\delta$-function, conductivity integral for $\omega>0$, and the
kinetic energy are sometimes rearranged in the form
\begin{equation}
 \begin{array}{l}
 \frac{\omega_{p,s}^2}{8}=
 \frac{a^2 \pi e^2}{2\hbar^2}\langle - H_{kin} \rangle
     -\int_{0^+}^{\omega_m}\mbox{Re}\sigma(\omega)d\omega
 \end{array}
 \label{fsum4b}
\end{equation}
Whenever the kinetic energy term on the right-hand side changes
its value, this expression suggests a 'violation' of the f-sum
rule, since the spectral weight in the $\delta$-function now no
longer compensates the change of spectral weight in the
conductivity integration on the right-hand side. Of course there is
no real violation, but part of the optical spectral weight is
being swapped between the intra-band transitions and the inter-band
transitions. Later in this chapter we will use the relation
between kinetic energy as expressed in the original incarnation
due to Maldague \cite{maldague77} (Eq. \ref{fsum4}) to determine
in detail the temperature dependence of the $ab$-plane kinetic
energy of some of the high T$_c$ superconductors.

It is easy to see, that for a small filling fraction of the band
Eq. \ref{fsum4} is the same as the Galilean invariant result: The
occupied electron states are now all located just above the bottom
of the band, with an energy $-t$. Hence in leading orders
of the filling fraction $-\langle\psi_g|H_{t}|\psi_g\rangle=Nt$.
Identifying $a^2\hbar^{-2}t^{-1}$ as the effective mass $m^*$ we
recognize the familiar f-sum rule, Eq.\ref{fsum}, with the free
electron mass replaced by the effective mass.

As the total spectral weight (intra-band plus inter-band) should
satisfy the f-sum rule, the intra-band spectral weight is
bounded from above, {\em i.e.}
$0 \le \sum_k n_k/m_k \le n/m$.
Near the top of the band the dynamical mass has the peculiar property
that it is negative, $m_k < 0$, which in the present
context adds a negative contribution to the intra-band spectral weight.
On the other hand, the fact that Re$\sigma(\omega)$ has to be larger than zero,
implies that the equilibrium momentum distribution function $n_k$ is
subject to certain bounds: If for example $n_k$ would preferentially
occupy states near the top of the band, leaving the states at the
bottom empty, the intra-band spectral weight would acquire an unphysical
negative value. Apparently such momentum distribution functions
cannot result from the interactions of Eq.\ref{hamiltonian},
regardless of the strength and k-dependence of those interactions.

\subsection{Additional sum rules for $\sigma(\omega)$ and $1/\epsilon(\omega)$}\label{sec:moresum}
Several other sum rule type expressions exist for the optical
conductivity and for the dielectric constant. Here we give a
summary. In the presence of a magnetic field an optical analogue
of the Hall effect exists. The behavior is similar to the
DC-limit, resulting in an off-diagonal component of the optical
conductivity $\sigma_{xy}(\omega)=-\sigma_{yx}(\omega)$, where the
$z$-axes is parallel to the magnetic field. The optical Hall angle
is defined as
\begin{equation}\label{opticalhallangle}
 t_H(\omega)=\frac{\sigma_{xy}(\omega)}{\sigma_{xx}(\omega)}
\end{equation}
The optical ($\sigma_{xx}$) and Hall conductivities($\sigma_{xy}$)
can be measured directly in optical transmission experiments
\cite{spielman,kaplan}. Drew and Coleman have shown\cite{drew97}
that this response function obeys the sum rule
\begin{equation}\label{hallsum}
 \frac{2}{\pi}\int_0^{\infty} t_H(\omega)d\omega = \omega_H
\end{equation}
where the Hall frequency $\omega_H$ is unaffected by interactions,
and in the Galilean invariant case corresponds to the bare
cyclotron frequency, $\omega_H=eB/m$.

A {\em first moment} sum rule of the optical conductivity is easily
obtained for $T=0$, by direct integration of Eq. \ref{sigmaA},
providing
\begin{equation}
  \begin{array}{l}
     \int_{0}^{\infty} \omega\sigma_{\alpha,\alpha}(q,\omega)d\omega =
     \frac{2\pi}{\hbar V} \langle j_{\alpha,q}j_{\alpha,-q} \rangle =\\
      = \frac {2\pi e^2 \hbar}{m^2 V}
      \sum_{k,\sigma,\sigma'} k_{\alpha}^2  \langle
      c^{\dagger}_{k-q/2,\sigma}c_{k+q/2,\sigma'}c^{\dagger}_{k+q/2,\sigma'}c_{k-q/2,\sigma}\rangle
  \end{array}
 \label{1tmomenta}
\end{equation}
In free space there is no scattering potential nor a
periodic potential causing Umklapp scattering. Hence for electrons moving
in free space the right-hand side of Eq. \ref{1tmomenta}
is exactly zero. This comes as no surprise: The integral on the left-hand
side is also zero, since the optical conductivity of such
a system has only a $\delta$-function at $\omega=0$ due to Galilean invariance.
However, in the presence of Umklapp scattering the eigen-states of the
electrons with energy-momentum dispersion $\epsilon_k$ are no longer the free electron
states in the summation of Eq. \ref{1tmomenta}. The true eigen-states are superpositions
of plane waves. Vice versa the free electron states generated by the $c^{\dagger}_k$ operators
of the above expression can be written as a superposition of the eigen-states of the
periodic potential: $c^{\dagger}_{k+G,\sigma} = \sum_{m} \alpha_G^m(k) a^{\dagger}_{k,m,\sigma}$,
where the latter operator generates the $m$'th eigen-state with momentum $k$ in the
first Brillouin zone. For brevity we introduce the notation $A_G^m = |\alpha_G^m(k)|^2$,
and $\hat{n}_{\sigma}^{j}=a^{\dagger}_{k,j,\sigma}a_{k,j,\sigma}$. Expressed in terms of these band
occupation number operators Eq. \ref{1tmomenta} is
\begin{equation}
     \lim_{q\rightarrow0}\int_{0}^{\infty} \omega\sigma_{\alpha,\alpha}(q,\omega)d\omega =
      \frac {2\pi e^2 \hbar}{m^2 V}
      \sum_{k,G} (k_{\alpha}+G_{\alpha})^2 \sum_{j,m,\sigma} A_G^j A_G^m
      \langle \hat{n}_{\sigma}^{j}(1-\hat{n}_{\sigma}^{m})\rangle
 \label{1tmomentb}
\end{equation}
The summation on the right-hand side strongly suggests an
intimate relationship between the optical conductivity and the
kinetic energy of the electrons. However, due to the fact that the
expression on the right-hand side is rather difficult to calculate,
the first moment of $\sigma(\omega)$ is of little practical
importance. It's main purpose in the present context is to
demonstrate the trend that an increase of the kinetic energy is
accompanied by an increase of the first moment of the optical
conductivity spectrum. This is consistent with the notion, that an
increase of kinetic energy is accompanied by a blue-shift of the
spectral weight.

For the energy-loss function a separate series of sum rule type equations can be
derived\cite{mahan81,nozieres66,turlakov02}
\begin{equation}
     \int_{-\infty}^{\infty} \mbox{Im}\frac{-\omega}{\epsilon(\omega)}d\omega =
     \frac{4\pi^2  e^2 N}{mV}
 \label{eps+1sum}
\end{equation}
which is similar to the f-sum rule for the optical conductivity, Eq. \ref{fsum}.

As a result of the fact that the real and imaginary part of the energy loss
function are connected via Kramers-Kronig relations, the following relation
exists
\begin{equation}
     \int_{-\infty}^{\infty} \mbox{Im}\frac{-1}{\omega\epsilon(\omega)}d\omega = \pi
 \label{eps-1sum}
\end{equation}
This expression can in principle be used to calibrate the absolute
intensity of an energy-loss spectrum, or to check the experimental
equipment, since the right-hand side does not depend on any
parameter of the material of which the spectrum is taken. We can
use the relation between $\epsilon(\omega)$ and $\sigma(\omega)$,
Eq. \ref{epsilonsigma}, to express Eq. \ref{eps-1sum} as a
function of $\sigma(\omega)$. Using Cauchy's theorem, it is quite
easy to prove from Eq.\ref{eps-1sum}, that
\begin{equation}
     \int_{0}^{\infty} \mbox{Re}
     \frac{1} {\sigma(\omega)-i\lambda \omega}d\omega
     = \frac{\pi}{2\lambda}
 \label{tausum1}
\end{equation}
Often the intra-band optical conductivity is analyzed in terms of a
frequency dependent scattering rate
$1/\tau(\omega)=(ne^2 / m) \mbox{Re}\{\sigma(\omega)^{-1}\}$,
which follows directly from the experimental real and imaginary
part of the optical conductivity. Taking the limit $\lambda\rightarrow 0$
in Eq.\ref{tausum1}, we observe that
\begin{equation}
     \int_{0}^{\infty}\frac{1} {\tau(\omega)}d\omega
     = \lim_{\lambda\rightarrow 0} \frac{\pi}{2\lambda} \frac{n e^2}{m} = \infty
 \label{tausum2}
\end{equation}
Hence ultraviolet divergency appears to be a burden of
integral formulas of the frequency dependent scattering
rate\cite{basov01b,shah01a,abanov01} which is hard to avoid.

In section \ref{sec:ecorr} we will encounter a relation between the
loss-function and the Coulomb energy stored in the electron fluid\cite{nozieres66}
\begin{equation}
 \int_{0}^{\infty} \mbox{Im}\frac{-1}{\epsilon(\vec{k},\omega)}d\omega =
 \frac{4\pi^2 e^2}{\hbar|\vec{k}|^2}
 \langle\Psi_0|\hat{\rho}_k\hat{\rho}_{-k} |\Psi_0 \rangle
 \label{coulombsum}
\end{equation}
This expression is limited to the ground state at $T=0$,
as was also the case for Eq. \ref{1tmomenta}.
The integrands on the left-hand side of
Eq. \ref{1tmomenta} and Eq. \ref{coulombsum}
are {\em odd} functions of frequency. In contrast
the f-sum rule, and the other expressions given in this subsection all
involve integrals over an even function of frequency, which is the
reason why the latter can be represented as integrals over all (positive and
negative) frequencies. The fact that $\hbar$ occurs on the right-hand side of
Eqs. \ref{1tmomenta} and \ref{coulombsum} implies that these expressions
are of a fundamental quantum mechanical nature, with no equivalent in
classical physics.

Recently Turlakov and Leggett derived an expression for the {\em third} moment
of the energy loss function, which in the limit of $k\rightarrow 0$ is a function
of the Umklapp potential of Eq.\ref{fullhamiltonian}
\begin{equation}
     \int_{-\infty}^{\infty} \mbox{Im}\frac{-\omega^3}{\epsilon_{\alpha\alpha}(\omega)}d\omega =
     \frac {4\pi^2}{m^2}  \left\langle -\sum_{G}G_{\alpha}^2U_G \hat{\rho}_{-G} \right\rangle
 \label{eps+3sum}
\end{equation}
The fact, that the right-hand side of Eq. \ref{eps+3sum} is finite
implies, that for $\omega\rightarrow\infty$ the loss function of
any substance must decay more rapidly than
$\mbox{Im}\{-\epsilon(\omega)^{-1}\} \propto \omega^{-4}$, and
that the optical conductivity decays faster than
$\mbox{Re}\{\sigma(\omega)\} \propto \omega^{-3}$. This expression
is potentially interesting for the measurement of changes in
Umklapp potential, provided that experimental data can be
collected up to sufficiently high photon energy, so that the left-hand
side of the expression reaches its high frequency limit.

\section{The internal energy of superconductors}
A necessary condition for the existence of superconductivity is, that
the free energy of the superconducting state is lower than that of the
non-superconducting state. At sufficiently high temperature
important contributions to the free energy are due to the entropy. These
contributions depend strongly on the nature of the low energy
excitations, first and foremost of all their nature be it fermionic,
bosonic or of a more complex character due to electron correlation
effects. At $T=0$ the free energy and internal energy are equal, and
are given by the quantum expectation value of the
Hamiltonian, which can be separated into
an interaction energy and a kinetic energy.
\subsection{Interaction energy in BCS theory}\label{sec:ecorr}
\begin{figure}[ht]
\centerline{\includegraphics[width=10cm,clip=true]{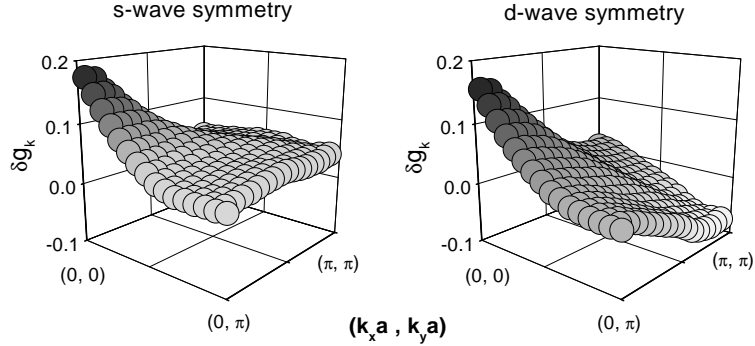}}
  \caption{\protect The k-space representation of the superconductivity induced change of
  pair correlation function
  for the s-wave (left panel) and d-wave symmetry (right panel).
  Parameters: $\Delta/W = 0.2$, $\omega_D/W=0.2$. Doping level x = 0.25}
\label{fig:gk}
\end{figure}
We consider a system of electrons interacting via the interaction
Hamiltonian given in Eq. \ref{hamiltonian}.
In the ground state of the system, the interaction energy, including the
correlation energy beyond the Hartree-Fock approximation,
is just the quantum expectation value of the second (interaction) term of \ref{hamiltonian}.
Here we are only interested in the difference in interaction
energy between the normal and superconducting state.
\begin{equation}
 E_{corr}^s - E_{corr}^n = \sum_k V_k \left(\langle\hat{\rho}_k\hat{\rho}_{-k}\rangle_s-
 \langle\hat{\rho}_k\hat{\rho}_{-k}\rangle_n\right)=\sum_k V_k\delta g_{k}
  \label{ecorr}
\end{equation}
In BCS theory the only terms of the interaction Hamiltonian which
contribute to the pairing are the so-called reduced terms, {\em
i.e.} those terms in the summation of Eq. \ref{hamiltonian} for
which the center of mass momentum $p+q=0$. The quantum mechanical
expectation value of the correlation function is
\begin{equation}
 \delta g_{k}  = \sum_p
  (|u_{p+k}|^2-\theta_{p+k})(\theta_{p}-|u_{p}|^2)+\sum_p u_{p+k}v_{p+k}u_{p}^*v_{p}^*
 \label{gk}
\end{equation}
The first term on the right-hand represents the change in
exchange correlations, whereas the second term represents the particle-hole
mixing which is characteristic for the BCS state.
A quantity of special interest is the real space correlation function
$\delta g(r,r') = \langle n(r)n(r')\rangle_s-\langle n(r)n(r')\rangle_n$.
The Fourier transform of this correlation function is directly related to
$\delta g_{k}$ appearing in the expression of the interaction energy, Eq.\ref{ecorr}
\begin{equation}
 \delta g_{k} =
 \frac{1}{V^2} \int d^3r \int d^3r' e^{ik(r-r')} \delta g(r,r')
 \label{corfun}
\end{equation}
We see, that if the correlation function $\delta g(r,r')$ could be
measured somehow, and the interaction $V_k$ is known, than
the interaction energy would follow directly from
our knowledge of $\delta g(r,r')$:
\begin{equation}
 E_{corr}^s-E_{corr}^n = \int d^3r \int d^3r' V(r-r') \delta g(r,r')
 \label{ecorrinr}
\end{equation}
In a conventional superconductor the quasi-particles of the normal
state are also the fermions which become paired in the
superconducting state. (Note, that now we are using the concept of
Landau Fermi-liquid quasi-particles for the normal state. Later in
this manuscript we will explore some consequences of {\em not}
having a Fermi liquid in the normal state, where the
quasi-particle concept will be abandoned.) Although the
quasi-particle eigen-states of a conventional Fermi liquid have an
amount of electron character different from zero, their effective
masses, velocities and scattering rates are renormalized. The
conventional point of view is, that pairing (enhancement of
pair correlations) reduces the interaction  energy of the
electrons, by virtue of the fact that in the superconducting state
the pair correlation function $g(r,r') =
\langle\Psi|\hat{n}(r)\hat{n}(r') |\Psi\rangle$ increases at
distances shorter than the superconducting coherence length
$\xi_0$. If the interaction energy $V(r-r')$ is {\em attractive}
for those distances, the interaction energy, Eq. \ref{ecorrinr},
decreases in the superconducting state, and $V(r-r')$ represents a
(or the) pairing mechanism.
\begin{figure}[th]
\centerline{\includegraphics[width=10cm,clip=true]{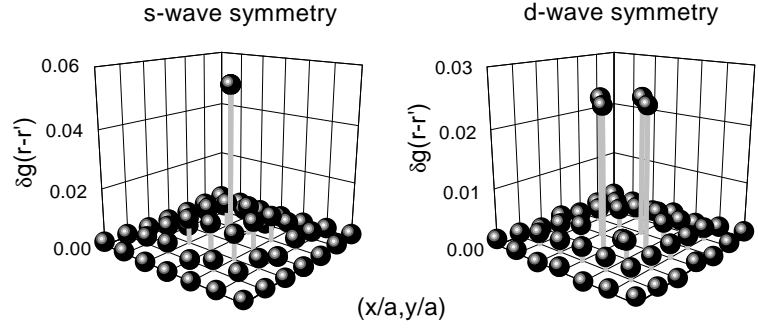}}
  \caption{\protect The coordinate space
  representation of the superconductivity induced change of
  pair correlation function
  for the s-wave (left panel) and d-wave symmetry (right panel).
  Parameters: $\Delta/W = 0.2$, $\omega_D/W=0.2$. Doping level:
  x = 0.25}
\label{fig:gr}
\end{figure}
In Fig. \ref{fig:gk} we show calculations of $\delta g_{k}$
assuming a bandstructure of the form
\begin{equation}
 \epsilon_{k} = \frac{W}{4}\left[ \cos k_xa+\cos k_ya \right]-\mu
\end{equation}
while adopting an order parameter of the form
\begin{equation}
 \Delta_{k} = \Delta_0 \Theta( |\epsilon_k-\mu|-\omega_D)
\end{equation}
for s-wave symmetry, and
\begin{equation}
 \Delta_{k} = \Delta_0\left[\cos k_xa-\cos k_ya \right]
 \Theta( |\epsilon_k-\mu|-\omega_D )
\end{equation}
for d-wave symmetry. The parameters used were $\Delta/W=0.2$, $\omega_D/W=0.2$,
and $E_F/W=0.43$ corresponding to x=0.25 hole doping counted from
half filling of the band. The chemical potential in the superconducting state
was calculated selfconsistently in order to keep the hole doping at the fixed value
of x=0.25 \cite{marel90,marel92,rietveld92,marel94}.
From Fig.\ref{fig:gk} we conclude that s-wave pairing symmetry requires a negative $V_k$
regardless of the value of $k$, whereas the d-wave symmetry can be stabilized either
assuming $V_k>0$ for $k$ in the $(\pi,\pi)$ region, or $V_k<0$ for $k$ near the
origin. Both types of symmetry are suppressed by
having $V_k > 0$ at small momentum, such as the Coulomb interaction.

In Fig. \ref{fig:gr} we display the correlation
function in coordinate space representation.
This graph demonstrates, that d-wave pairing is stabilized by a nearest-neighbor
attractive interaction potential. An on-site {\em repulsion} has no influence
on the pairing energy, since the pair correlation function has zero amplitude
for $r-r'=0$. On the other hand, for s-wave pairing the 'best' interaction
is an on-site attractive potential, since the s-wave $\delta g(r,r')$ reaches it's maximum
value at $r-r'=0$.

\subsection{Experimental measurements of the Coulomb interaction energy\label{expcoul}}
In a series of papers Leggett has discussed the change of Coulomb
correlation energy for a system which becomes superconducting\cite{ajl99},
and has argued, that this energy would actually decrease
in the superconducting state.
Experimentally the changes of Coulomb energy can be measured
directly in the sector of $k$-space of vanishing $k$. The best,
and most stable, experimental technique is to measure the
dielectric function using spectroscopic ellipsometry, and to
follow the changes as a function of temperature carefully as a
function of temperature. Because the cuprates are strongly
anisotropic materials, it is crucial to measure both the in-plane
and out-of-plane pseudo-dielectric functions, from which the full
dielectric tensor elements along the optical axes of the crystal
then have to be calculated. We followed this procedure for a
number of different high T$_c$ cuprates, indicating that the
Coulomb energy in the superconducting state {\em in}creases for
k=0. However, for $k\neq0$ this need no longer be the case.
Summarizing the situation\cite{thesispresura}: the Coulomb interaction energy
increases in the superconducting state for small $k$. This
implies, that the lowering of internal energy in the
superconducting state must be caused either by other sectors of
k-space (in particular at around the $(\pi,\pi)$ point, see Fig.
\ref{fig:gk}!), or by a lowering of the kinetic energy in the
superconducting state. The latter is only possible in a non Fermi
liquid scenario of the normal state.

\subsection{Kinetic energy in BCS theory}\label{sec:ekin}
\begin{figure}[ht]
\centerline{\includegraphics[width=10cm,clip=true]{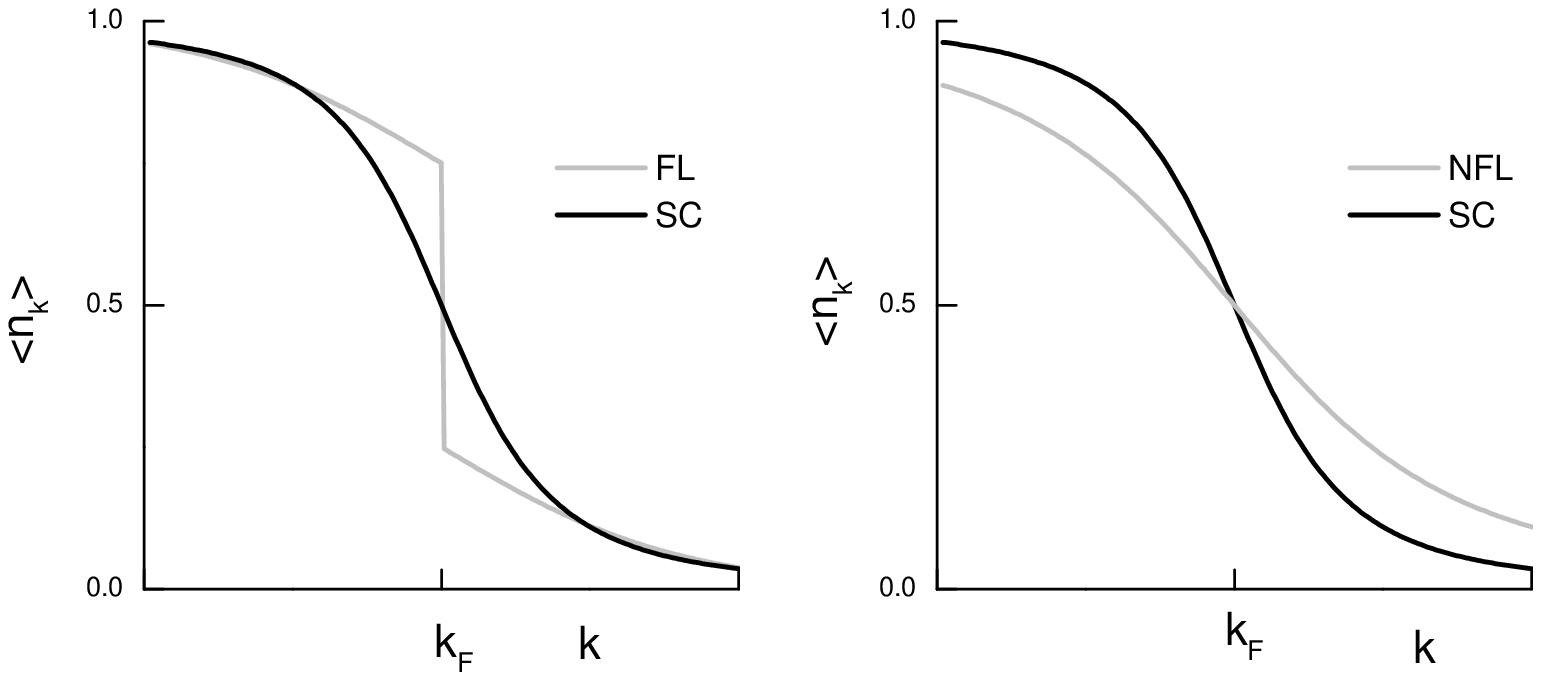}}
  \caption{\protect Occupation function as a function of momentum in the normal (dash)
  and the superconducting (solid) state for Fermi liquid (left panel)
  and an example of a broad distribution function, not corresponding to a
  Fermi liquid (right panel).}
\label{fig:nk}
\end{figure}
In BCS theory the lowering of the pair-interaction energy is
partly compensated by a change of kinetic energy of opposite sign.
This can be understood qualitatively in the following way: The
correlated motion in pairs causes a localization of the relative
coordinates of electrons, thereby increasing the relative momentum
and the kinetic energy of the electrons. Another way to see this,
is that in the superconducting state the step of $n_k$ at the
Fermi momentum is smoothed, as indicated in the left panel of Fig.
\ref{fig:nk}, causing $E_{kin}$ to become larger\cite{norman00}.

A pedagogical example where
the kinetic energy of a pair is higher in the superconducting state, is
provided by the negative $U$ Hubbard model\cite{micnas90}: Without interactions, the
kinetic energy is provided by the expression
\begin{equation}
 E_{kin} =  -t \sum_{<i,j>,\sigma}\langle\Psi|c^{\dagger}_{i\sigma}c_{j\sigma}+H.c.|\Psi\rangle
\end{equation}
Let us consider a 2D square lattice. If the band contains two electrons, the kinetic
energy of each electron is $-2t$, the bottom of the band, hence
$E_{kin}=-4t$. (In a tight-binding picture the reference energy is the center of the band
irrespective of $E_F$, causing $E_{kin}$ to be always negative). Let us now consider
the kinetic energy of a pair in the extreme pairing limit, {\em i.e.} $U\gg t$, causing
both electrons to occupy the same site, with an interaction energy $-U$. The occupation
function $n_k$ in this case
becomes
\begin{equation}
 n_k \approx \frac{1}{N_k}\frac{t}{U}\frac{1}{(1+4\epsilon_k/U)^2}
\end{equation}
This implies
that the kinetic energy approaches $E_{kin} \rightarrow -8t^2/U$.
Hence the kinetic energy increases from
$E_{kin}^n = -4t$ to $E_{kin}^s =  -\frac{8t^2}{U}$ when the local pairs are formed.
The paired electrons behave like bosons of charge $2e$. A second order perturbation
calculation yields an effective boson hopping
parameter\cite{nozieres85} $t'=t^2/U$. In experiments probing the charge dynamics, this
hopping parameter determines the inertia of the charges in an accelerating field. As
a result the plasma frequency of such a model would be
\begin{equation}
 \omega_{p,s}^2= 4\pi \frac{n}{2} (2e)^2 \frac{a^2 t^2}{\hbar^2U}
 \label{localpair}
\end{equation}
whereas if these pair correlations are muted
\begin{equation}
 \omega_{p,n}^2= 4\pi n e^2 \frac{a^2 t}{\hbar^2}
\end{equation}
Because the plasma frequency is just the low frequency spectral
weight associated with the charge carriers, this
demonstrates, that for conventional pairs ({\em i.e.} those which
are formed due to interaction energy lowering) the expected trend
is, that in the superconducting state the spectral weight {\em
de}creases. Note, that this argument can only demonstrate the direction in which
the plasma frequency changes when the pair correlations become reduced, but
it does not correctly provide the quantitative size of the change, since
the strong coupling regime of Eq. \ref{localpair} implies the
presence of a finite fraction of uncondensed
'preformed' pairs in the normal state.
The same effect exists in the limit of weak pairing correlations.
In Ref. \cite{marel95} (Eq. 29, ignoring particle-hole asymmetric
terms) the following expression was derived for the plasma
resonance
\begin{eqnarray}
  \omega_{p,s}^2 = \frac{4\pi e^2}{V}\sum_k\frac{\Delta_k^2}{\hbar^2E_k^3}
  \left[\frac{\partial\epsilon_k}{\partial k}\right]^2
\end{eqnarray}
where $V$ is the volume of the system, and $E_k^2=\epsilon_k^2+|\Delta_k|^2$.
Integrating in parts, using that
$\Delta_k^2 E_k^{-3} \partial_k \epsilon_k = \partial_k\left(\epsilon_k/E_k \right)$,
and that $\partial_k\epsilon_k=0$ at the zone-boundary, we obtain
\begin{equation}
 \omega_{p,s}^2 = \frac{4\pi e^2}{V} \sum_k \frac{n_k}{m_k}
 \label{wps}
\end{equation}
where $m_k^{-1} = \hbar^{-2}\partial^2\epsilon_k/\partial k^2$,
and $n_k=1-\epsilon_k/E_k$.
For a monotonous band dispersion the
plasma frequency of the superconductor is always
{\em smaller} than that of the unpaired system:
Because the sign of the band-mass changes from positive near the bottom of the band to
negative near the top, the effect of the broadened occupation factors $n_k$ is to
give a slightly smaller average over $m_k^{-1}$, hence $\omega_p^2$ is smaller. Note that
the mass of free electrons does not depend on momentum, hence in free space $\omega_p^2$ is
unaffected by the pairing.

To obtain an estimate of the order of magnitude of the change of spectral weight, we
consider a square band of width W with a Fermi energy $E_F=N_e/(2W)$, where
$N_e$ is the number of electrons per unit cell. To simplify matters we assume that
$1/m_k$ varies linearly as a function of band energy:$1/m(\epsilon)=(W-2E_F-2\epsilon)/(Wm_0)$.
We consider the limit where $\Delta << W,E_F$.
Let us assume that the bandwidth $\sim 1$ eV, and $\Delta \sim 14$ meV corresponding
to $T_c=$90 K. The reduction
of the spectral weight is then 0.28 $\%$. If we assume that the bandwidth is 0.1 eV,
the spectral weight reduction would typically be 11.4 $\%$.

\subsection{Kinetic energy driven superconductivity}\label{sec:ekindriven}
If the state above T$_c$ is {\em not} a Fermi liquid, the
situation could be reversed. The right-hand panel of
Fig.\ref{fig:nk} represents a state very different from a
Fermi liquid, and in fact looks similar
to a gapped state. Indeed even for the 1D Luttinger liquid $n(k)$
has an infinite slope at $k_F$. If indeed the normal state
would have a broad momentum distribution like the one indicated, the
total kinetic energy becomes lower once pairs are formed, provided that
the slope of $n(k)$ at $k_F$ is steeper in the superconducting
state. This is not
necessarily in contradiction with the virial theorem, even though
ultimately all relevant interactions (including
electron-phonon interactions) are derived from the Coulomb
interaction: The superconducting correlations involve the low
energy scale quasi-particle excitations and their interactions.
These {\em effective} interactions usually have characteristics
quite different from the original Coulomb interaction, resulting
in $E_c/E_{kin}\neq -2$ for the low energy quasi-particles.
Various models have been recently proposed involving pairing due
to a {\em reduction} of kinetic energy. In strongly anisotropic
materials such as the cuprates, two possible types of kinetic
energy should be distinguished: Perpendicular to the
planes\cite{pwa95,chakra98} (along the c-direction) and along the
planar directions\cite{hirsch92a,hirsch92b,chakravarty93,
alexandrov94,emery95,assaad96,lee99,pwa00,emery00}.

\section{Experimental studies of superconductivity induced spectral weight transfer}
\label{sec:ekinopt}
\subsection{Josephson plasmons and c-axis kinetic energy}
\label{sec:ekinperp}
\begin{figure}[ht]
\centerline{\includegraphics[width=10cm,clip=true]{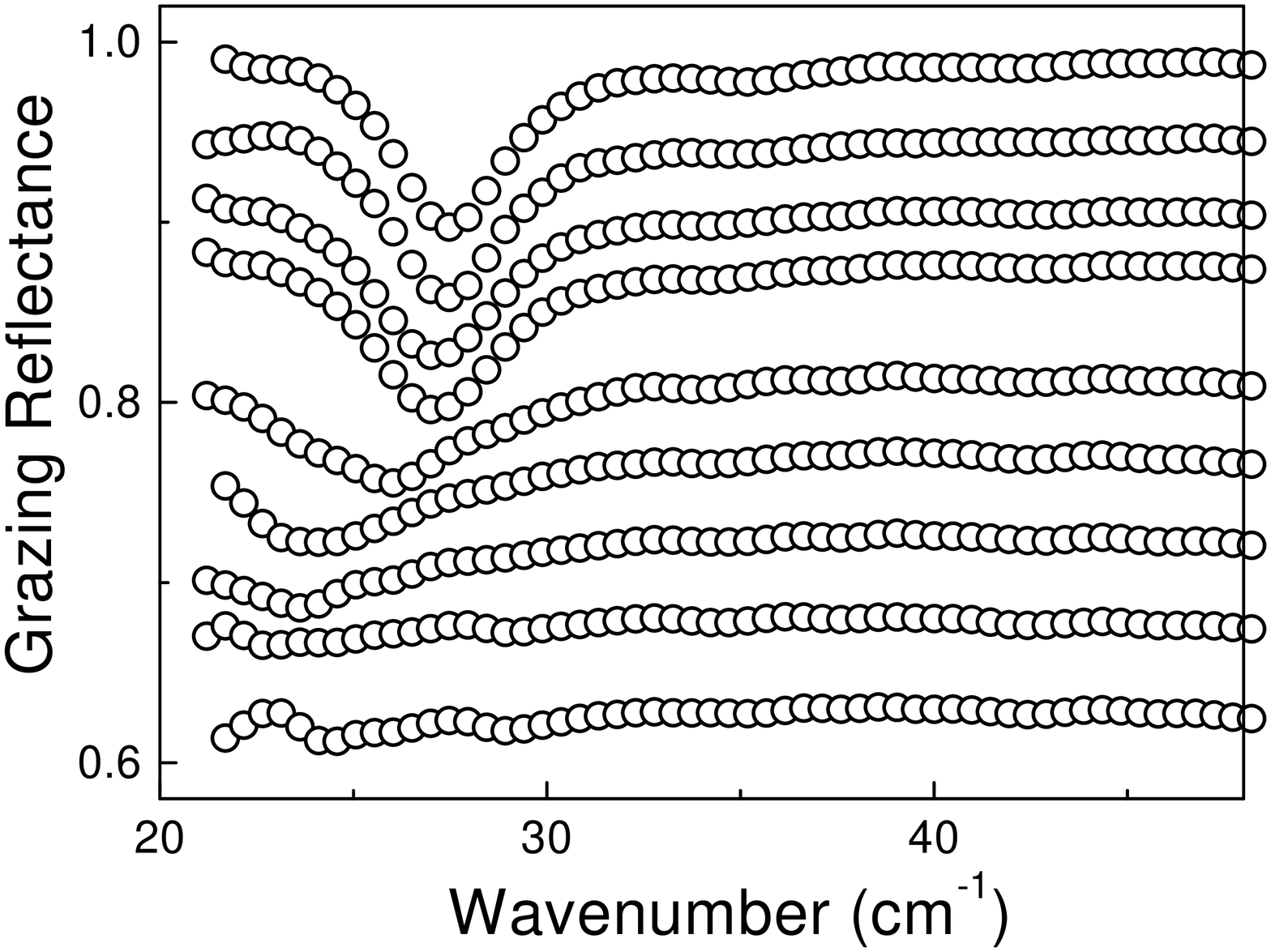}}
  \caption{\protect
  P-polarized reflectivity at 80$^o$ angle of incidence of Tl$_2$Ba$_2$CuO$_6$.
  From top to bottom: 4K, 10 K, 20 K, 30 K, 40 K, 50 K, 60 K, 75 K,
  and 90 K. The curves have been given incremental 3 percent vertical offsets for clarity.
  Data from Ref. \cite{tsvetkov98}}
\label{fig:tl2201}
\end{figure}

\begin{figure}[th]
\centerline{\includegraphics[width=10cm,clip=true]{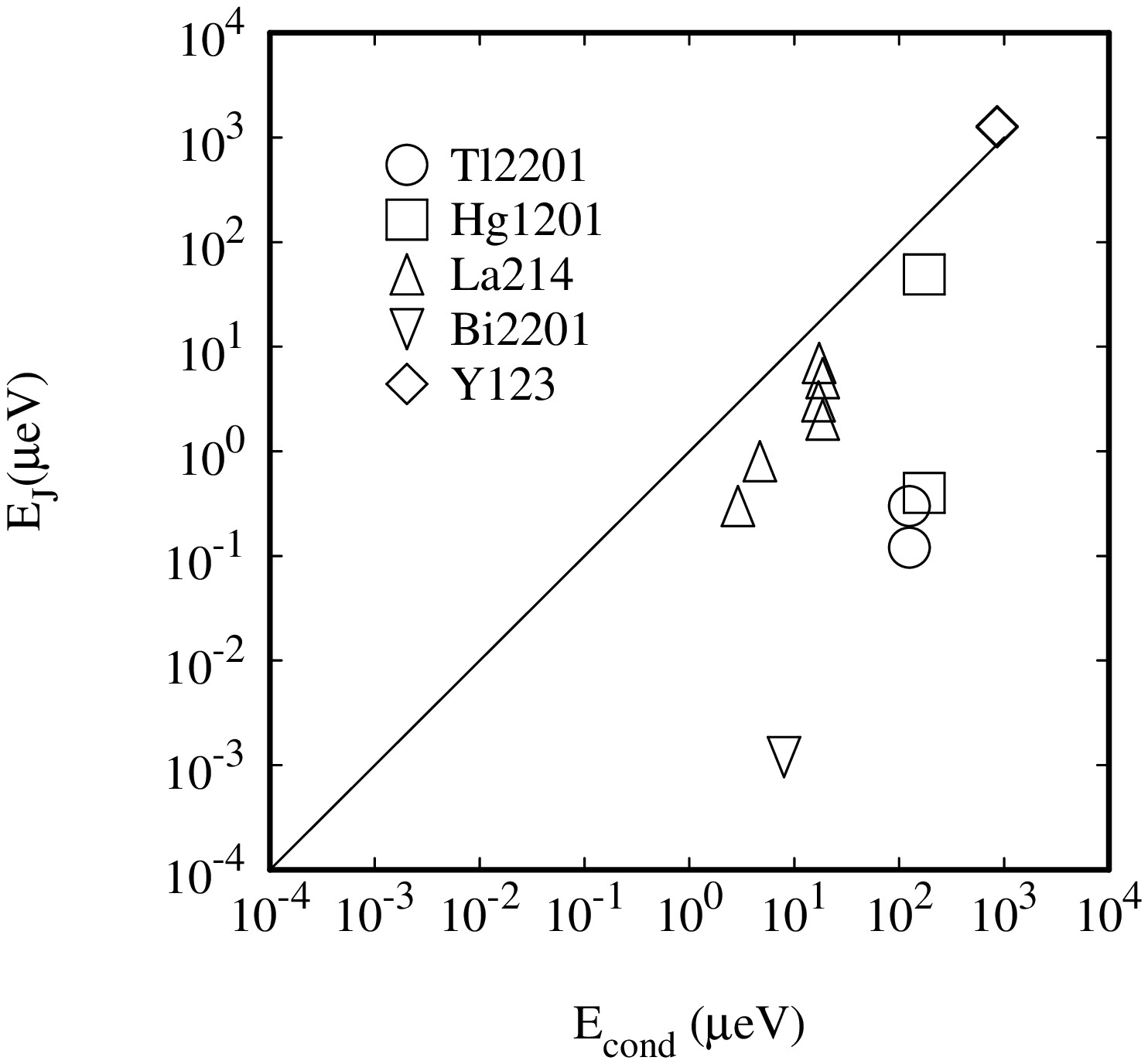}}
  \caption{\protect Intrinsic Josephson coupling energy
  \cite{basov99,panagop97,tsvetkov98,kam98,kirtley99,gaifullin99,kim94,tamasaku92,grueninger00}
  versus condensation energy\cite{loram94,loram01}}
\label{fig:econd}
\end{figure}

C-axis kinetic energy driven superconductivity has been proposed
within the context of inter-layer tunneling, and has been
extensively discussed in a large number of
papers\cite{hirsch92a,hirsch92b,pwa95,ajl96,chakra98,chakravarty99,marel96a,
schuetzmann97,panagop97,tsvetkov98,kam98,basov99,kirtley99,gaifullin99,basov01a,zelezeny01,munzar01,
munzar02,boris02}. One of the main reasons to suspect that
superconductivity was c-axis kinetic driven, was the observation
of "incoherent" c-axis transport of quasi-particles in the normal
state\cite{cooper94} and, rather surprisingly, {\em also} in the
superconducting state\cite{kim94,dulic99,hosseini98}, thus
providing a channel for kinetic energy lowering for charge
carriers as soon as pairing sets in. As discussed in
section \ref{sec:ksum} a very useful tool in the discussion of
kinetic energy is the low frequency spectral weight associated
with the charge carriers. In infrared spectra this spectral
weight is contained within a the 'Drude' conductivity peak
centered at $\omega=0$. Within the context of the tight-binding
model a simple relation exists between the kinetic energy per
site, with volume per site $V_u$, and the low frequency spectral
weight\cite{maldague77,baeriswyl86}
\begin{equation}
 E_{kin} = \frac{\hbar^2 V_u}{4\pi e^2a^2} \omega_p^2
 \label{tbfsumB}
\end{equation}
Here the plasma frequency, $\omega_p$, is used to quantify the low frequency
spectral weight:
\begin{equation}
\frac{\omega_{p,s}^2}{8} + \int_{0^+}^{\omega_m}
\mbox{Re}\sigma(\omega) d\omega = \frac{1}{8}\omega_p^2
 \label{fA}
\end{equation}
where the integration should be carried out over all
transitions within the band, {\em including} the
$\delta$-function at $\omega=0$ in the superconducting state.
The $\delta(\omega)$ peak in Re$\sigma(\omega)$ is of course not
visible in the spectra directly. However the presence of the
superfluid is manifested prominently in the London term of
Re$\epsilon(\omega)$  (proportional to Im$\sigma(\omega)$):
$\epsilon_L(\omega)=-\omega_{p,s}^2\omega^{-2}$. In La$_{2-x}$Sr$_x$CuO$_4$
the London term is manifested in a spectacular way as a prominent plasma
resonance perpendicular to the superconducting planes\cite{tamasaku92}.
This is commonly used to determine the superfluid spectral weight,
$\omega_{p,s}^2$, from the experimental spectra. Apart from
universal prefactors, the amount of spectral weight of the
$\delta(\omega)$ conductivity peak corresponds to the Josephson
coupling energy, which in turn is the inter-layer {\em
pair}hopping amplitude. It therefore provides an upper limit to
the change of kinetic energy between the normal and
superconducting state\cite{pwa95,ajl96}, because the spectral
weight transferred from higher frequencies to the
$\delta(\omega)$-peak cannot exceed this amount. This allowed a
simple experimental way to test the idea of c-axis kinetic energy
driven superconductivity by comparing the experimentally measured
values of the condensation energy ($E_{cond}$) and E$_J$. The inter-layer tunneling
hypothesis required, that $E_J\approx E_{cond}$. In the spring of
1996 the first experimental results were presented\cite{marel96a}
for Tl2201 (Tc=80 K), showing that $E_J$ was at least two orders
of magnitude too small to account for the condensation energy.
\begin{figure}[ht]
\centerline{\includegraphics[width=10cm,clip=true]{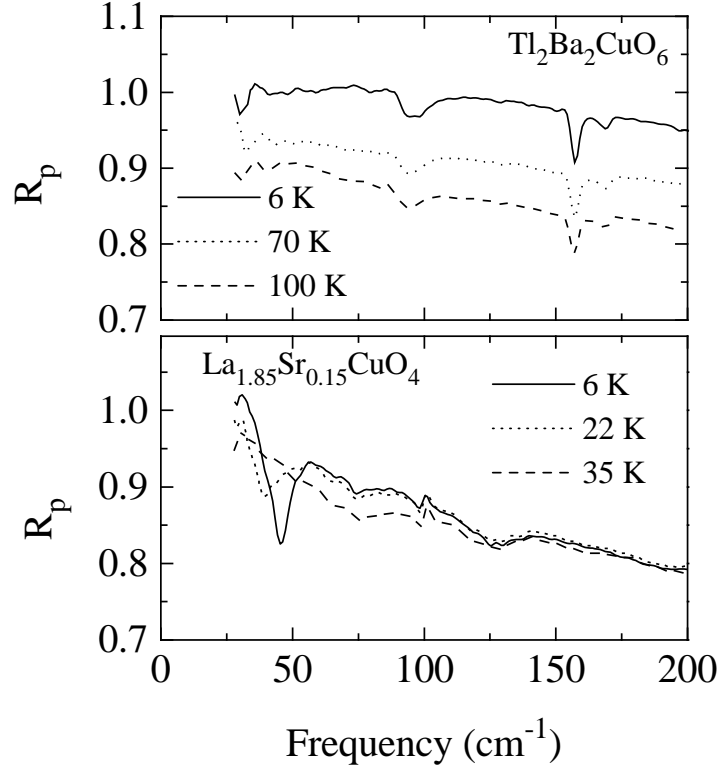}}
  \caption{\protect Grazing reflectivity of a Tl$_2$Ba$_2$CuO$_6$
  thin film (upper panel) and La$_{1.85}$Sr$_{0.15}$CuO$_4$
  single crystal (lower panel) measured with the polarization
  of the incident light tilted at an angle of 80$^\circ$ relative to
  the copper-oxygen planes. For LSCO the Josephson plasma resonance can be
  clearly seen at 40 cm$^{-1}$. For Tl2212 no the Josephson plasma resonance
  is observed, indicating that it is located below the lower limit of
  30 cm$^{-1}$ of the spectrometer. This implies that the Josephson
  coupling energy in this compound is at least two orders of magnitude
  lower than required by the inter-layer tunneling hypothesis.
  Data from Ref. \cite{schuetzmann97}}
\label{fig:tlfig3}
\end{figure}
Later measurements of $\lambda_c$\cite{kam98} (approximately 17 $\mu$m) and
the Josephson plasma resonance (JPR)\cite{tsvetkov98} at $28$ cm$^{-1}$,
allowed a definite determination of the Josephson coupling energy of this
compound, indicating that $E_J\approx 0.3 \mu$eV
in Tl2201 with $T_c=80$ K (see Fig. \ref{fig:tl2201}).
This is a factor 400 lower than $E_{cond}\approx 100 \mu$eV per
copper, based either on $c_V$ experimental data\cite{loram94}, or on the formula
$E_{cond}=0.5N(0)\Delta^2$ with $N(0)=1 eV^{-1}$ per copper, and $\Delta\simeq 15 meV$.
In Fig. \ref{fig:econd} the change in c-axis kinetic energy and the Josephson
coupling energies are compared to the condensation energy for a large number
of high T$_c$ cuprates. For most materials we see that
$E_J < E_{cond}$, sometimes differing by several orders of magnitude.

These arguments falsifying the inter-layer tunneling mechanism have been questioned\cite{chakravarty99}, arguing
that a large part of the specific heat of Tl2201 is due to 3D fluctuations, and that these fluctuations should
be subtracted when the condensation energy is calculated. However, it was recently shown\cite{marel02} that due
to thermodynamical constraints the fluctuation correction can not exceed a factor 2.5 in the case of Tl2201 (as
compared to a factor 40 in Ref. \cite{chakravarty99}). Hence the discrepancy between the Josephson coupling
energy and the condensation energy of Tl2201 is still two orders of magnitude.

However, as stressed above, $E_J$ provides only an {\em upper
limit} for $\Delta E_{kin}$. A c-axis kinetic energy change {\em
smaller} than $E_J$ is obtained if we take into account the fact
that a substantial part of $\delta(\omega)$-function is just the
spectral weight removed from the sub-gap region of the optical
conductivity. Usually it is believed that in fact the latter is
the {\em only} source of intensity of spectral weight for the
$\delta$-function, known as the (phenomenological)
Glover-Tinkham-Ferell\cite{tinkham59} sum rule. According to the
arguments given in section \ref{sec:ekindriven} we may conclude
that $E_{kin,s}=E_{kin,n}$ when we observe, that {\em all}
spectral weight origins from the far-infrared gap region in
agreement with the Glover-Tinkham-Ferrell sum rule. If, on the
other hand, superconductivity is accompanied by a lowering of
c-axis kinetic energy, part of $\omega_{p,s}^2$ originates from
the higher frequency region of inter-band transitions, which begins
at typically 2 eV. In other words, we may say that
$\omega_{p,s}^2$ is an upper limit to the kinetic energy change
\begin{equation}
 0 < E_{kin,n}-E_{kin,s} < \frac{\hbar^2V_u}{4\pi e^2a^2} \omega_{p,s}^2
 \label{swsum}
\end{equation}
A direct determination of $E_{kin,s}-E_{kin,n}$ is obtained by
measuring experimentally the amount of spectral weight transferred
to the $\delta(\omega)$ peak due to the passage from the normal to
the superconducting state, as was done by Basov {\em et
al.}\cite{basov99,basov01a}. These data indicated that for
under-doped materials about 60$\%$ comes from the sub-gap region in
the far infrared, while about 40$\%$ originates from frequencies
much higher than the gap, whereas for optimally doped cuprates at
least 90$\%$ originates from the gap-region, while less than
10$\%$ comes from higher energy. Experimental artifacts caused by
a very small amount of mixing of ab-plane reflectivity into the
c-axis reflectivity curves may have resulted in an overestimation
of the spectral weight originating from high
energies\cite{basov01a}, in particular those samples where the
electronic $\sigma_c(\omega)$ is very low due to the
2-dimensionality. Optimally doped YBCO is probably less prone to
systematic errors due to leakage of $R_{ab}$ into the c-axis
reflectivity, since $\sigma_c(\omega)$ of this material is among
the largest in the cuprate family. The larger $\sigma_c(\omega)$
causes the c-axis reflectivity to be much larger at all
frequencies, thereby reducing the effect of spurious mixing of
$ab$-plane reflectivity in the optical spectra on the
Kramers-Kronig analyzes.

In summary $\Delta E_{kin,c} < 0.1 E_J$ in most
cases. For several of the single-layer cuprates it has become clear now, that
$\Delta E_{kin}$ significantly undershoots the condensation energy,
sometimes by two orders of magnitude or worse, as indicated in Fig. \ref{fig:econd}.

\subsection{Josephson plasmons in multi-layered cuprates}\label{sec:josephson}

\begin{figure}[t]
\centerline{\includegraphics[width=10cm,clip=true]{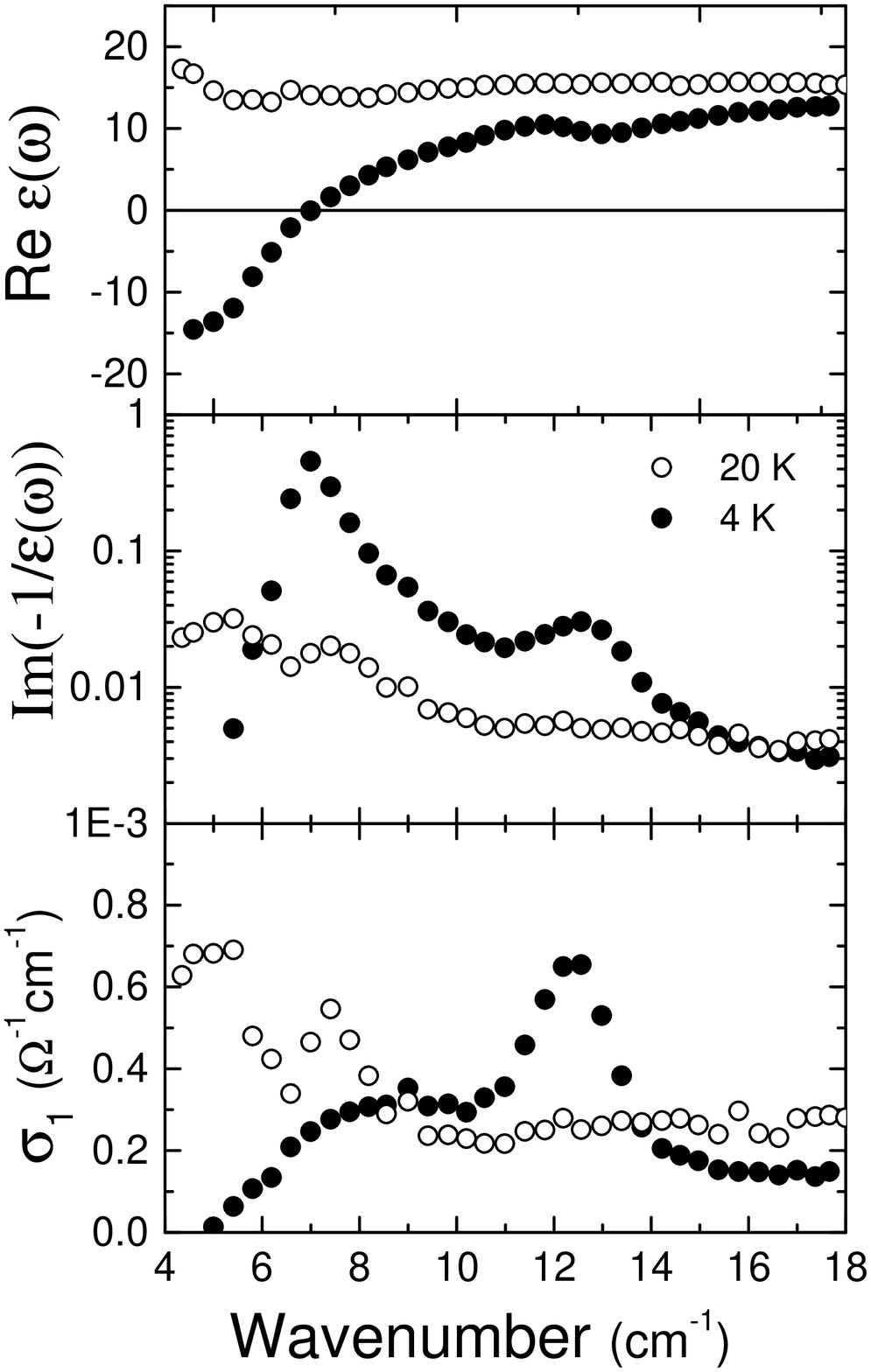}}
  \caption{(a) Real part of the $c$-axis dielectric function
   of SmLa$_{0.8}$Sr$_{0.2}$CuO$_{4-\delta}$ for 4 K (closed symbols), and 20 K (open symbols)
   (b) The c-axis loss function, Im$\epsilon(\omega)^{-1}$.
   (c) Real part of the $c$-axis optical conductivity.
   Data from Ref. \cite{dulic01,thesisdiana}.}
 \label{fig:slsco}
\end{figure}

This situation may be different for the bi-layer compounds. In
these materials in principle the coupling within the bi-layer may
provide an additional source of frustrated inter-layer kinetic
energy, which can in principle be released when the material
enters the superconductng state. This can in principle be
monitored with infrared spectroscopy, because quite generally a
stack of Josephson coupled layers with two different types of weak
links alternating (in the present context corresponding to
inter-bilayer and intra-bilayer) should exhibit three Josephson
collective modes instead of one: Two of those modes are
longitudinal Josephson plasma resonances, which show up as peaks
in the energy loss function Im$(-1/\epsilon(\omega))$. In between
these two longitudinal resonances one expects a transverse optical
plasma resonance, which is revealed by a peak in
Re$\sigma(\omega)$. In essence the extra two modes are
out-of-phase oscillations of the two types of junctions. This has
been predicted in Ref. \cite{marel96b} for the case of a
multi-layer of Josephson coupled 2D superconducting layers.
Further detailed calculations for the bi-layer case were presented
in Refs. \cite{marel01,shah01b}. The existence of {\em two}
longitudinal modes and {\em one} associated transverse plasmon
mode at finite frequencies has been confirmed experimentally for
the SmLa$_{0.8}$Sr$_{0.2}$CuO$_{4-\delta}$ in a series of
papers\cite{shibata98,shibata01,kakeshita01,dulic01,pimenov01}
(see Fig. \ref{fig:slsco}).

\begin{figure}[th]
\centerline{\includegraphics[width=10cm,clip=true]{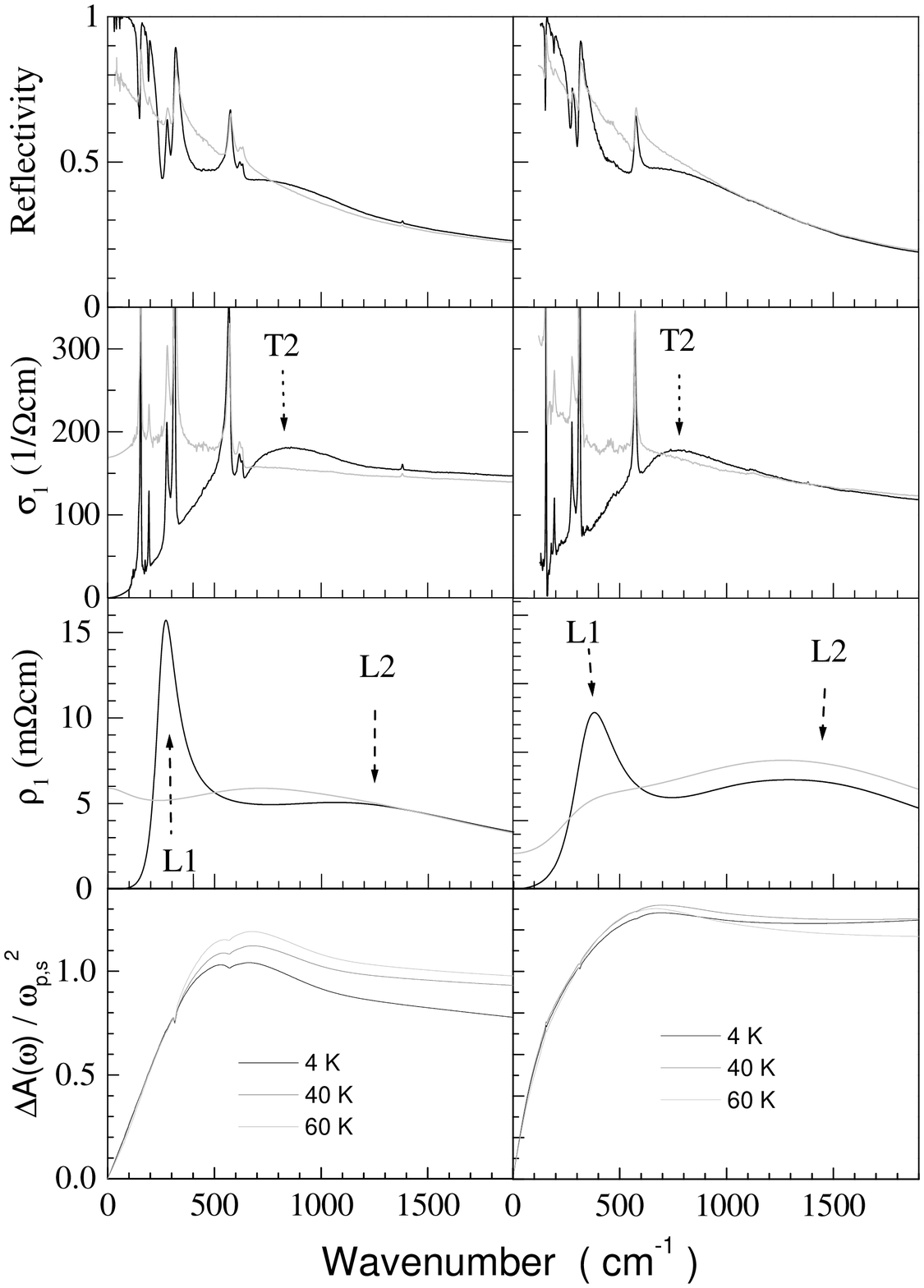}}
  \caption{\protect C-axis optical spectra of optimally doped (x=6.93)
  and over-doped (x=7.0) YBa$_2$Cu$_3$O$_{7-x}$.
  From top to bottom: reflectivity, optical conductivity, dynamical impedance and
  relative spectral weight (Eq. \ref{eq:relweight}). The dynamical impedance,
  $\rho_1(\omega)$=Re$4\pi/\omega\epsilon(\omega)$ is proportional to the energy loss
  function weighted by a factor $1/\omega$. The optical phonons have
  been subtracted from the loss-functions for clarity. The data are from
  Ref. \cite{grueninger00,thesismarkus}}
\label{fig:ybcopt}
\end{figure}

The c-axis optical conductivity of YBCO is one order of magnitude
larger than for LSCO near optimal doping. As a result the relative
importance of the optical phonons in the spectra is diminished. In
the case of optimally doped YBCO, the experiments indicate no
significant transfer of spectral weight from high frequencies
associated with the onset of superconductivity. C-axis
reflectivity data\cite{grueninger00} of optimally doped YBCO are
shown in Figs.\ref{fig:ybcopt}. Above T$_c$ the optical
conductivity is weakly frequency dependent, and does not resemble
a Drude peak. Below T$_c$ the conductivity is depleted for
frequencies below 500 cm$^{-1}$, reminiscent of the opening of a
large gap, but not an s-wave gap, since a relatively large
conductivity remains in this range.

There is a slight overshoot of the optical conductivity
in the region between 500 and 700 cm$^{-1}$, due to the fact that the
normal state and superconducting state curves cross at 600 cm$^{-1}$. In the
case of the bi-layer cuprates this could be explained as a result of the presence
of two superconducting layers per unit cell, resulting in the
'transverse optical' plasma mode mentioned
above\cite{munzar01,munzar02,boris02,munzar99,zelezny99,grueninger00}.

For the f-sum rule the presence of this extra mode makes no
difference. The extra spectral weight in the superconducting state
associated with this mode has in principle the same origin as the
spectral weight in the zero-frequency $\delta$-function. In a
conventional picture the source would be the spectral weight,
removed due to a depletion of $\sigma_c(\omega)$ in the
gap-region. The implementation of the sum rule relevant for this
case then states that the relative spectral weight function
\begin{equation}
 \frac{\Delta A(\omega)}{\omega_{p,s}^{2}} =
 \frac{8}{\omega_{p,s}^{2}}\int_{0^+}^\omega \left(\sigma_n(\omega')-\sigma_s(\omega')\right)d\omega'
 \label{eq:relweight}
\end{equation}
overshoots the 100 $\%$ line close to the 'second plasma' mode, and saturates at
100 $\%$ for frequencies far above this mode. This is indeed observed in
Fig. \ref{fig:ybcopt}.

Additional studies of the bi-layer (and tri-layer) materials have
provided confirmation of the transverse optical plasmon in these
materials. In spite of its high frequency, making the assignment
to the Josephson effect rather dubious, nevertheless the
transverse optical mode either makes its first appearance below
T$_c$, or gains in sharpness and intensity at the temperature where
pairs are being formed (which for under-doped cuprates begins
already above T$_c$). Also in at least a number of cases the
spectral weight of the 'transverse optical' plasmon observed below
T$_c$ appears to originate not from the spectral weight removed
from the gap region, but from much higher
energies\cite{basov01a,zelezeny01,munzar01,munzar02,boris02}. The
implication of this may be, that a non-negligible fraction of
frustrated c-axis kinetic energy is released when these materials
become superconducting. This seems to be particularly relevant for
the strong intra-bilayer (or tri-layer) coupling of Bi2212, Bi2223
and Y123.

\subsection{Kinetic energy parallel to the planes}\label{sec:ekinpar}
In-plane kinetic energy driven superconductivity has been proposed
by a number of researchers:
Hirsch\cite{hirsch92a,hirsch92b,hirsch98} discussed this
possibility as a consequence of particle-hole asymmetry. It has
also been discussed within the context of holes moving in an
anti-ferromagnetic
background\cite{bonca89,riera89,hasagewa89,dagotto90,barnes92,demler98}.
More recently the possibility of a {\em reduction} of kinetic
energy associated with pair hopping between stripes has been
suggested\cite{emery95,emery00}, and an in-plane
pair delocalization mechanism have been proposed in the context of
the resonating valence band model\cite{lee99,pwa00}.

A major issue is the question how to measure this. The logical
approach would be to measure again $\sigma(\omega,T)$ using the
combination of reflectivity and Kramers-Kronig analysis, and then
compare the spectral weight function in the superconducting state
to the same above T$_c$. There are several weak points to this
type of analysis. In the first place there is the problem of
sensitivity and progression of experimental errors: Let us assume,
that the change of kinetic energy is of order 0.1 meV per Cu atom
(this is approximately the condensation energy of the optimally
doped single layer cuprate Tl2201, with T$_c$ = 85 K.). For an
inter-layer spacing of 1.2 nm, this corresponds to a spectral
weight change $\Delta (\nu_p^2) = 10^5$ cm$^{-2}$. As the total
spectral weight in the far infrared range is of order
$\nu_p^2=14000^2$cm$^{-2}$, the relative change in spectral weight
is of order 0.05 $\%$. Typical accuracy reached for spectral
weight estimates using conventional reflection techniques is of
order 5$\%$. This illustrates the technical difficulties one has
to face when attempting to extract superconductivity induced
changes of the kinetic energy.

Experimental limitations on the accuracy are imposed by (i) the
impossibility to measure {\em all} frequencies including the
sub-mm range, (ii) systematic errors induced by Kramers-Kronig
analysis: The usual procedure is to use data into the visible/ultra-violet
range and beyond for completing the Kramers-Kronig analysis in the far infrared,
assuming that no important temperature dependence is present
outside the far infrared range. Obviously this assumption becomes
highly suspicious if the search is concentrated on spectral weight
transfer originating from precisely this frequency range.

The remedy is, to let nature perform the spectral weight integral. Due to causality
Re$\epsilon(\omega)$ and Re$\sigma(\omega)$ satisfy the Kramers-Kronig
relation
\begin{equation}
 \mbox{Re}\epsilon(\omega)=1-\int_0^{\infty}\frac{8\mbox{Re}\sigma(z)}{\omega^2-z^2}dz
 \label{kkr}
\end{equation}
The main idea of spectral weight transfer is, that spectral weight is essentially transferred from
the inter-band transitions at an energy of several eV, down to the $\delta$-function
in $\sigma(\omega)$ at $\omega=0$. Indeed various groups have reported a change of
optical properties in the visible part of the spectrum when the sample becomes
superconducting\cite{holcomb96,little99,ruebhausen01,molegraaf02}.
If this is the case, we have $x=0$ for
the extra spectral weight in relation \ref{kkr}.
Together with Eq. \ref{swsum} it follows that changes in kinetic energy can be
read directly from Re$\epsilon(\omega)$ using the relation
\begin{equation}
 \delta E_{kin}^{eff}(\omega) = \frac{4\hbar^2\omega^2 V_u}{\pi e^2 a^2} \mbox{Re} \delta\epsilon(\omega)
\end{equation}
If the spectral weight is transferred to a frequency range $\omega_0$, then the above
expression can still be applied for $\omega \gg \omega_0$. If we measure Re$\epsilon(\omega)$
{\em directly} using spectroscopic ellipsometry, then indeed  nature does the
integration of $\sigma(\omega)$ for us at each temperature. This eliminates to
a large extent various systematic errors affecting the overall
accuracy of the spectral weight sum. It is important to measure
the complex dielectric constant for a large range of different
frequencies.

The second problem is that already above the superconducting phase transition the
optical spectra of these materials have appreciable temperature dependence.
What we really like to measure is the spectra of the same material in the superconducting
state, and in the 'normal' state, both at the same temperature. Typical magnetic fields
required to bring the material in the normal state are impractical, let alone the
complications of magneto-optics which then have to be faced. A more practical approach
is to measure carefully the temperature dependence over a large temperature range, with
small temperature intercepts, and to search for changes which occur at the phase transition.

\begin{figure}[th]
\centerline{\includegraphics[width=10cm,clip=true]{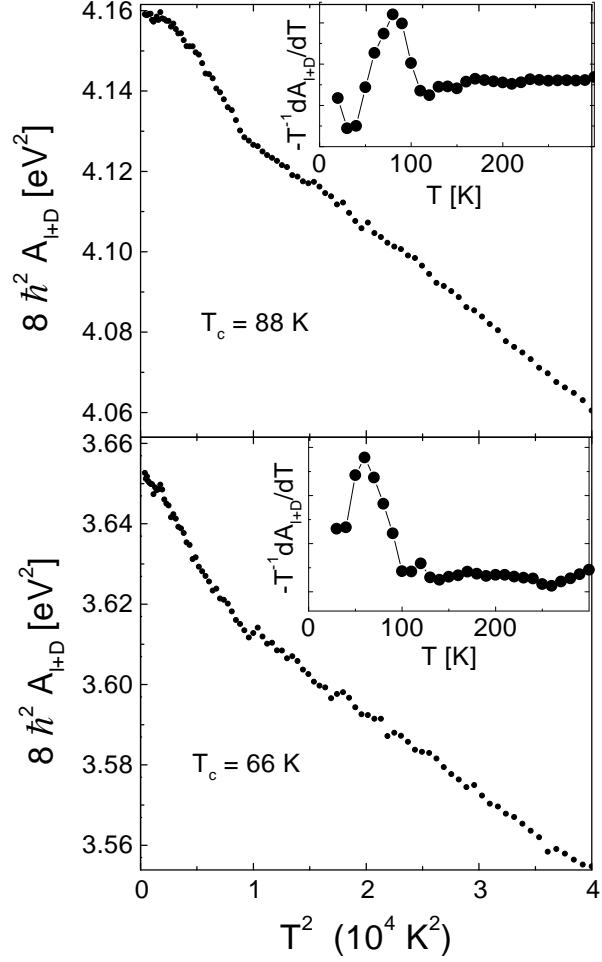}}
  \caption{\protect Measured values of the quantity
  $c^2\omega_{p,s}^2+8\int_{0^+}^{\infty}\mbox{Re}\sigma_{ab}(\omega)d\omega$
  of Bi2212 (T$_c$=88 K). The data are taken from
  Ref. \cite{molegraaf02,santander02}. To make the conversion to kinetic
  energy summed over the two $ab$-plane directions, the numbers
  along the vertical axis have to be multiplied
  with a factor $-10^3V_u/(4\pi e^2 a^2) = -83$ meV / eV$^2$.
  }
\label{fig:kbi}
\end{figure}
In Fig. \ref{fig:kbi} the spectral weight from 0 to 10000
cm$^{-1}$ is shown as a function of temperature for the case of
Bi2212\cite{molegraaf02}. Note that this integral corresponds to
{\em minus} the ab-plane kinetic energy. We observe, that in the
superconducting state the kinetic energy drops by an amount of
about 1 meV per Cu. This is in fact a relatively large effect.
This surprising result seems to tell us that in the cuprates the
kinetic energy in the superconducting state is {\em lowered}
relative to the normal state. This corresponds to the
unconventional scenario depicted in the right-hand panel of
Fig.\ref{fig:nk}, where the normal state is a non Fermi liquid,
whereas the superconducting state follows the behavior of a (more)
conventional BCS type wave-function with the usual type of
Bogoliubov quasi-particles. The amazing conclusion from this would
be, that there is no need for a {\em lowering} of the
interaction energy any more. The condensation energy of optimally
doped Bi2212 is about 0.1 meV per Cu atom\cite{loram01}.
\section{Conclusions}
The optical conductivity is a fundamental property
of solids, contains contributions of vibrational and electronic character.
Among the electronic type of excitations the
intra-band and inter-band transitions, excitons,
and plasmons of different types correspond to the most prominent
features in the spectra. In addition multi-magnon excitations or more exotic
collective modes can often be detected. The careful study of the optical properties
of solids can provide valuable microscopic information about the electronic
structure of solids. In contrast to many other spectroscopic techniques, it
is relative easy to obtain reliable absolute values of the optical conductivity.
As a result sum rules and sum rule related integral expressions can often
be applied to the optical spectra. Here we have treated a few examples of
sum rule analysis: Application of the f-sum rule to the phonon spectra
of transition metal silicides provides information on the resonant electron-phonon
coupling in these materials. Integration of the energy-loss function gives the
value of the Coulomb energy stored in the material, which is seen to increase
when a high T$_c$ cuprate enters the superconducting state. The
spectral weight within the partially band of
the high T$_c$ cuprates is seen to become larger in superconducting
state. This effect exists both perpendicular to the planes and parallel
to the planes. This spectral weight change can be associated with a
decrease of kinetic energy when the material becomes superconducting. Although
the relative spectral weight change along the ab-plane is quite small, it
indicates a fairly large change of the ab-plane kinetic energy, large enough
to account for the energy by which the superconducting state of these materials
is stabilized.
In addition the real and imaginary part of the optical conductivity can be used to
study the intrinsic Josephson coupling between the superconducting planes. In
superconductors with two or more different types of weak links alternating, such
as SmLa$_{1-x}$Sr$_x$Cu$_4$, YBa$_2$Cu$_3$O$_{7-x}$, Bi$_2$Sr$_2$CaCu$_2$O$_8$, a rich
spectrum of plasma-oscillations is observed in the superconducting state, and sometimes
above $T_c$, due to the multi-layered structure of these materials. This has
provided important insights in the nature of the coupling, and it has been
used to extract quantitative values of this coupling.
\section{Acknowledgements}
Part of the experiments described in this chapter have been
carried out by J. H. Kim,
J. Schuetzmann, A. Tsvetkov, A. Kuzmenko,
G. Rietveld, B. J. Feenstra, H. S. Somal,
J. E. van der Eb, A. Damascelli, M. U. Grueninger, D. Dulic, C.
Presura, P. Mena, and H. J. A. Molegraaf. The author has
benefitted from discussions with, among others, D. I.
Khomskii, G. A. Sawatzky, M. J. Rice, P. W. Anderson, A. J.
Leggett, Z. X. Shen, S. C. Zhang, A. J. Millis, M. Turlakov, J. E.
Hirsch, D. N. Basov, M. R. Norman, and W. Hanke.

\begin{thebibliography}{1}
\bibitem{notepi}
        Note the extra factor of $\pi$. In Ref. \cite{nozieres58} this factor is absent,
        but  it is recuperated in the final formulas.
\bibitem{footnotesuperconductor}
        The intra-band optical spectrum of a superconductor
        typically has three contributions: (i) A Dirac-function in the
        origin corresponding to the supercondensate the presence of which is revealed by
        a reactive response in the imaginary part of the optical conductivity below the gap
        diverging as $\sigma_2(\omega)=c^2/(4\pi\lambda(T)^2\omega)$. (ii) A gapped Drude-like peak due
        absorption of light by pairs of quasi-particles excited above the gap. For $\omega >> 2\Delta$ the
        spectrum merges with the Drude peak observed in the normal state. (iii) A narrow peak of finite
        width centered at $\omega=0$ due to the dissipative response of thermal quasi-particles above the gap.
\bibitem{proof1} The proof of Eq. \ref{sum1} follows the same lines as the
        derivation of the f-sum rule. It employs the fact, that the current operators
        can be represented as commutators of the hamiltonian with the dipole operator.
        The expectation value of the hamiltonian is used to cancel out the factor $\omega_{mn}$
        in the denominator of the expression. In the final step the
        commutator of the dipole operator and the current operator is calculated in
        second quantization, which completes the derivation of Eq. \ref{sum1}.
        For a Galilean invariant system at $T=0$ this summation cannot be carried
        out directly for $q=0$. The singular behaviour at this point can be
        circumvented by considering the limit $q\rightarrow 0$.
\end{thebibliography}
\end{document}